%% file: APC_results_OM2023.tex
\documentclass[11pt]{article}
\usepackage[resetfonts]{cmap}
\usepackage{fancyvrb}
\usepackage{tabularray}
\usepackage{float}
\usepackage{graphicx}
\usepackage{codehigh}
\usepackage{subcaption}
\usepackage[normalem]{ulem}
\UseTblrLibrary{booktabs}
\UseTblrLibrary{siunitx}

\NewTableCommand{\tinytableDefineColor}[3]{\definecolor{#1}{#2}{#3}}
\usepackage{color}
\usepackage[rightcaption]{sidecap}
\usepackage{amsmath}
\usepackage{amsthm}
\usepackage{latexsym}
\usepackage[english]{babel}
\usepackage[margin= 1in]{geometry}
\usepackage{comment}
\usepackage{amssymb}
\usepackage{mathtools}
\usepackage{booktabs,array,dcolumn}
\usepackage{graphicx} 
\usepackage[unicode=true, bookmarks=true, bookmarksnumbered=false, bookmarksopen=false, breaklinks=true, pdfborder={0 0 0}, colorlinks = true, linkcolor=blue, citecolor=blue, urlcolor=blue, filecolor=blue] {hyperref}
\newcolumntype{d}{D{.}{.}{2.3}}
\newcolumntype{C}{>{\centering}p}
\usepackage{mathrsfs}
\usepackage{ulem}
\usepackage{nicefrac}
\usepackage[doublespacing]{setspace}
\usepackage{float}

\newtheorem*{Mech*}{Provider Side Mechanism}
\newtheorem*{lemma*}{Lemma}

\theoremstyle{definition}

\usepackage{tabularx}
\usepackage{natbib}
\usepackage{threeparttable}
\usepackage[group-separator={,}, group-minimum-digits=4]{siunitx}  
\usepackage{catchfile}                      
\newcommand*\inputNum[2][\num]{%
	\CatchFileDef{\tempnum}{#2}{}%
	#1{\tempnum}%
}
\usepackage{comment}                        
\definecolor{c1}{rgb}{0,0,1} 
\hypersetup{
	citecolor={c1}, 
}
\date{}
\begin{document}

\author{Pallavi Pal\thanks{Address: School of Business, Stevens Institute of Technology, NJ 07030, USA. Email: \href{mailto:ppal2@stevens.edu} {\tt ppal2@stevens.edu}.  I am grateful for the useful comments by Jay Pil Choi, Thomas Jietschko, Andrey Fradkin, and seminar participants at ISMS Marketing Science Conference 2021 for all the useful suggestions. Preliminary work, please do not share without permission.}}

\title{To Combine or Not? Consolidating  Horizontal Acquisitions in Multi-sided Market
 } 

\maketitle
\begin{abstract}

When a parent company acquires a horizontal competitor on the same side of a multi-sided market, it must decide whether to fully integrate the acquired platform or keep it as a separate brand. We study this in the context of Uber’s acquisition of Postmates, using novel consumer receipt data that tracks food delivery spending. Employing an Age–Period–Cohort (APC) decomposition, we isolate the merger’s effect on consumer spending while controlling for lifecycle and cohort effects. We find that Postmates users sharply reduced their spending on the platform after the merger, but spending shifted not only to UberEats, but also to competitors like DoorDash and Grubhub. Consumers who used multiple platforms and had low pre-merger activity on Postmates were more ``sticky," showing little change. Comparing our APC results with a standard Difference-in-Differences (DiD) design, we find the DiD underestimates the merger’s total impact by missing market-wide effects. Our findings suggest that in multi-sided markets, keeping acquired platforms separate can be beneficial; dissolving them may push demand to competitors, and some sticky multihoming users may not shift spending at all.\\
\vspace{2cm}
\end{abstract}

\newpage

\section{Introduction}\label{sec:intro}

The digital economy is increasingly characterized by parent companies that own multiple horizontal platforms competing on the same side of a multi-sided market. Meta operates both Facebook and Instagram as separate social media services; Alphabet maintains Google Search alongside YouTube; and in the food delivery sector, Uber acquired Postmates while continuing to operate UberEats. A central question in multi-sided market is whether such horizontal acquisitions lead to full platform integration or sustained coexistence---and how consumers respond to the resulting market restructuring. Understanding these dynamics and how the multi-sided nature changes the decision of how to optimally integrate firms. 

In this paper, we study the consumer-level consequences of Uber's acquisition of Postmates, which closed in December 2020. At the time, UberEats announced plans to integrate the platforms, which remain pending.\footnote{For details, see \hyperlink{https://help.uber.com/en/ubereats/restaurants/article/postmates-faqs?nodeId=063283d6-2602-490d-89b7-c9c15a07cfce}{here}} We examine whether Postmates users shifted spending to UberEats, or if the merger disruption redirected demand to competitors like DoorDash and Grubhub. We use a novel panel dataset of consumer receipts tracking spending across all major food delivery platforms before and after the merger.

Our empirical strategy employs an Age--Period--Cohort (APC) decomposition following \citet{oblander2023frontiers}. The key insight of this approach is that consumer spending patterns on a platform reflect three distinct forces: \emph{age effects} (how spending evolves with tenure on the platform), \emph{cohort effects} (systematic differences across users who joined at different times), and \emph{period effects} (time-varying shocks that hit all users simultaneously, such as a merger). By estimating these three components from a cohort $\times$ month panel and then forecasting what the period effects would have been absent the merger using an ARIMA counterfactual, we isolate the causal merger effect as the structural break in the period component.

We find a statistically significant decline of $-40.33$ percentage points in the Postmates budget share following the merger. However, the reallocation of spending was not directed solely toward UberEats, which gained only $+19.79$ pp. Our decomposition of other platform outcomes reveals that DoorDash ($+7.50$ pp) and Grubhub ($+10.49$ pp) captured a substantial fraction of the diverted Postmates spending, suggesting that the merger created competitive opportunities for rival platforms rather than consolidating demand within the Uber ecosystem.

Heterogeneity analysis provides further insight into which consumers drove these patterns. Splitting the sample by pre-merger Postmates budget share, we find that high-dependence users experienced the most pronounced declines ($-56.17$ pp), while low-share users saw a smaller shift of $-14.72$ pp. When we further interact budget share with MSA-level UberEats market concentration, the effect is sharpest among high-share single-homing users in high-UberEats markets ($-88.18$ pp)---consistent with the acquisition having the largest bite where the acquirer platform was most prominent. In contrast, consumers who had low budget allocated to Postmates experienced a more modest decline ($-14$ pp) and proved more ``sticky'': their budget allocation was less affected by the merger, suggesting that casual, diversified users had less reason to adjust their behavior.

We further estimate a Difference-in-Differences (DiD) framework that exploits cross-market variation in merger exposure. Specifically, the DiD design defines treatment based on whether an MSA had above-median pre-merger UberEats market share and estimates the differential effect on Postmates spending in treated versus control markets. The DiD specification curve remains stable across eight progressively controlled specifications. In addition, the heterogeneous treatment effect in DiD shows that users with a high Postmates budget share in treated markets experienced the largest declines. The results from DiD are comparable in direction and magnitude to the heterogeneous effect result in the main APC model. However, by construction, the DiD captures only the differential effect across markets and cannot recover the merger's level effect, which is common across markets. The APC decomposition complements the DiD by measuring the total merger-induced shift in spending, revealing that the overall effect is larger than what the cross-market comparison alone would suggest. This finding is consistent with the merger affecting consumer behavior through platform-wide channels (e.g., branding changes, app integration, supply-side consolidation) that operate independently of local market structure.

Our paper contributes to several strands of the literature. First, we contribute to the study of mergers and acquisitions in multi-sided platform markets. While a growing theoretical literature examines how platform mergers affect competition and welfare, and recent work explores merger simulation in digital markets \citep{kawaguchi2021merger}, empirical evidence on the consumer-side consequences of horizontal platform acquisitions remains scarce.\footnote{More broadly, our work relates to a growing empirical literature on network effects and competition in platform markets. Foundational theoretical models of multi-sided platforms \citep{rochet2003platform, armstrong2006competition, weyl2010price} predict that merging platforms can increase user value through network effects, but that platform competition may nonetheless benefit consumers through product variety and differentiation. Empirical tests of these predictions remain limited: \citet{rysman2004competition} finds positive cross-side network effects in Yellow Pages, while \citet{dube2010tipping} document market tipping in video game consoles. In the context of platform mergers, \citet{li2020measuring} exploit the combination of two ride-hailing platforms to estimate network effects, and \citet{cullen_outsourcing_2020} study local market structure in online services. } \citet{farronato2020dog} study network effects in a digital platform merger using a difference-in-differences design and similar to our results find limited consumer gains from consolidation. We compliment their work by looking at the market with competitors beyond the two merging platforms.  \citet{reshef2019smaller} examines how platform entry affects incumbents in online food delivery. \cite{mccarthy2023did} provide descriptive evidence on effect of multiple mergers among platforms. We provide individual-level evidence showing that the intended demand consolidation may fail when consumers have easy access to substitute platforms---a finding with direct implications for merger review in digital markets. Second, we contribute to the empirical M\&A literature in the food delivery industry. Generalizing empirical results in multi-sided markets is challenging because each market differs in the strength of network effects and degree of platform differentiation. Our analysis uses granular consumer receipt data that tracks individual spending across all major platforms, allowing us to trace exactly where displaced Postmates spending migrated after the merger. Furthermore, we provide evidence on how multihoming might effect the decision to consolidate merged horizontal platforms. Multi-homing behavior, a central feature of food delivery markets, has received growing theoretical attention \citep{bakos2019multihoming}, though empirical evidence on how multi-homing mediates the effects of horizontal acquisitions is scarce. Our paper provides direct evidence on this margin for a M\&A point of view.
Third, we demonstrate the value of Age--Period--Cohort decomposition methods---originally developed in marketing and demography \citep{oblander2023frontiers, blanchard2025game}---for identifying causal effects of discrete corporate events in panel data. By benchmarking the APC estimates against a standard DiD design, we show that the two approaches are complementary: the DiD captures differential effects across markets while the APC recovers the total merger-induced shift in consumer behavior.

The remainder of this paper is organized as follows. Section~\ref{sec:method} describes the APC methodology and identification strategy. Section~\ref{sec:results} presents the main event study results, heterogeneity analysis, and comparison with DiD. Robustness checks are provided in the appendix.

\section{Data}\label{section:Data}
We obtained a proprietary data set containing user-level transaction data which is supplemented with covid variables such as work from home order and covid cases across MSA. 


\subsection{Consumer Receipts Data} \label{subsection: receipts_data}
\begin{figure}[!htbp]
	\centering
	\caption{Display of apps}
	\includegraphics[height=8cm, width=15cm]{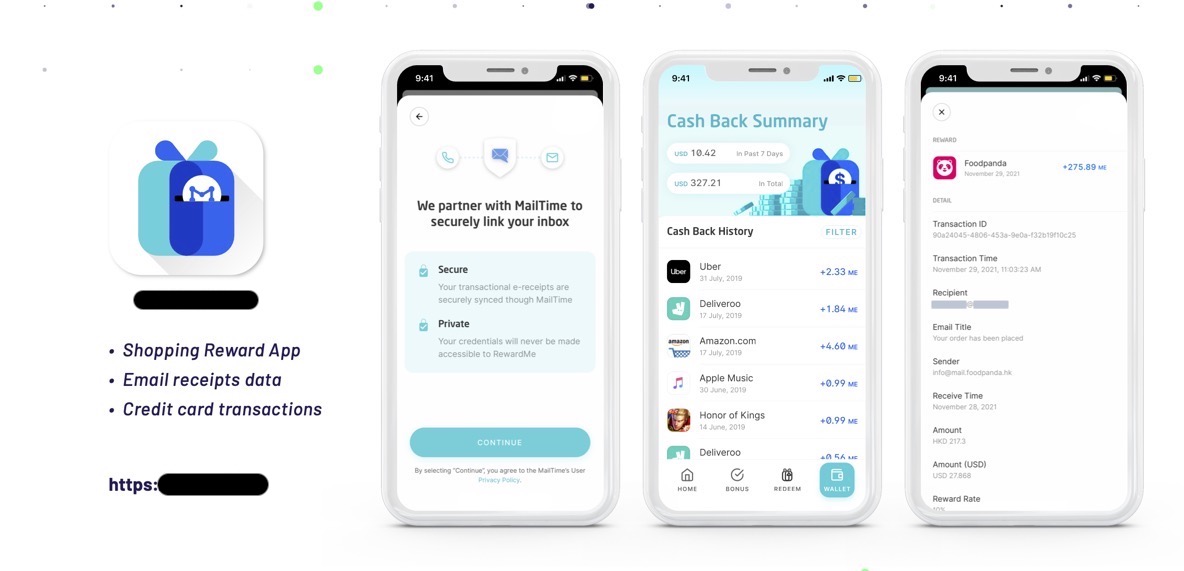}	\\
	\includegraphics[height=8cm, width=15cm]{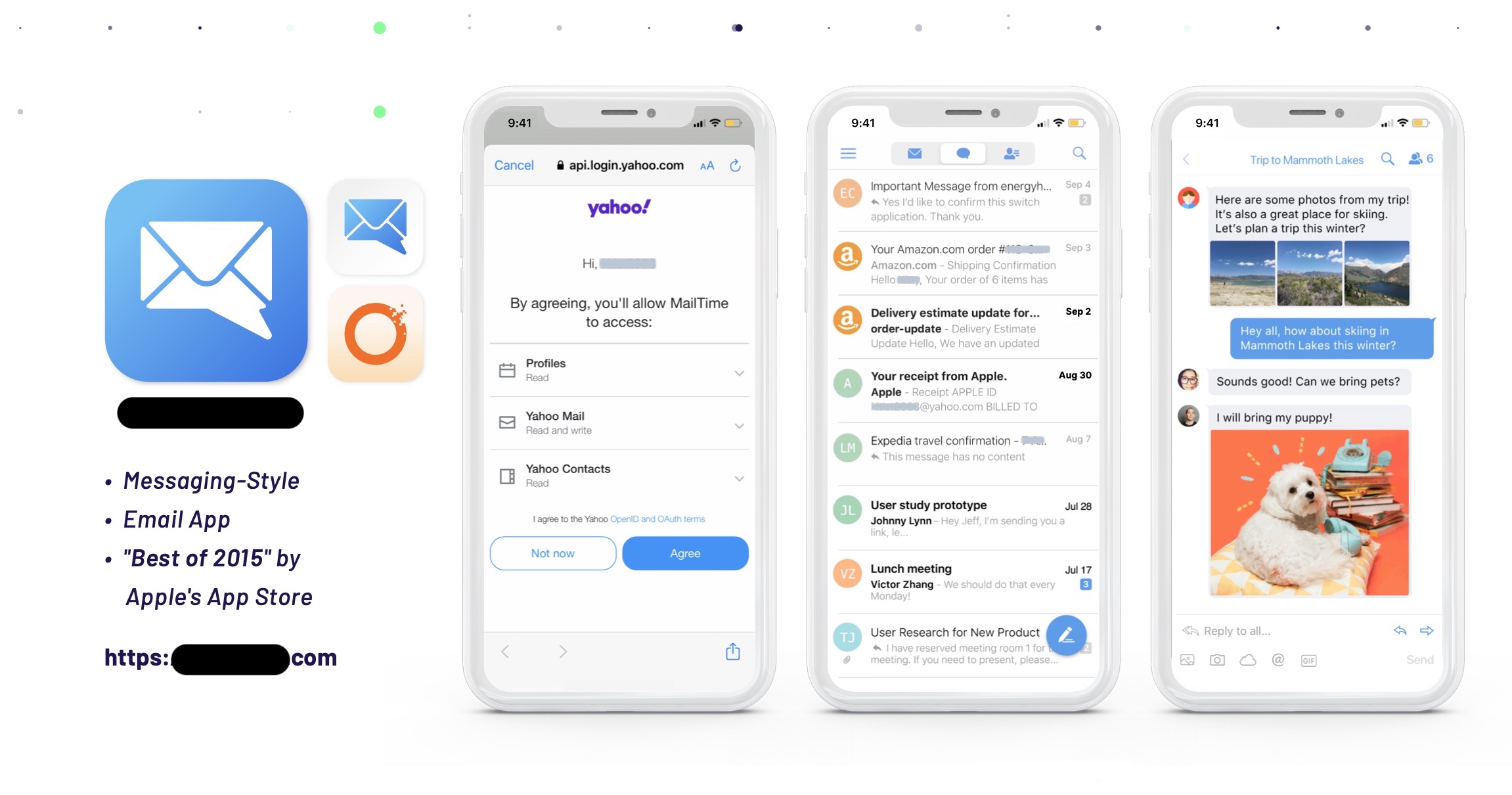}
	\label{fig: apps}
\end{figure}

We acquired proprietary data of the users from two lifestyle email apps that records all the email receipts of consumer orders from the four major food delivery apps from June 2019 to April 2022. Figure \ref{fig: apps} show display of the two apps. 


There are a total of \inputNum{tex/receipt_nreceipts.tex} receipts from \inputNum{tex/receipt_nusers.tex} active users in the dataset. Table \ref{tab: summary_stat} shows the summary statistics for the number of active users, the number of orders, and the average amount paid by each platform. There are fewest users on Postmates but making the highest average amount paid per order. 

\begin{table}[!htbp]
	\centering
	\caption{Summary Statistics by Platform}
	\input{tabs/receipts_summarystats.tex}
	\label{tab: summary_stat}
\end{table}

\subsubsection{Consumer Compositions on Platforms} \label{subsubsection: consumer_composition}

The receipts data contain consumers food delivery orders on four platforms: Doordash, Grubhub, Postmates, and UberEats. There are variations in consumer compositions on these platforms across geographic areas and over time. 

To further control for user specific spending habits, we concentrate on looking at percentage of total budget spent across platform each month. 

\subsection{Sample Construction}\label{subsec:sample}

Our analysis focuses on consumers who used Postmates prior to the merger announcement in December 2020. We construct the analysis sample through the following steps. First, we restrict to users with at least one Postmates transaction before December 2020 in a metropolitan statistical area (MSA) containing at least 10 such users. For users appearing in multiple MSAs, we assign them to their primary MSA based on the highest number of total orders. We then apply two data quality filters: (i) we remove users whose pre-merger Postmates budget share is below 5\%, as these represent negligible Postmates engagement; and (ii) we remove users who had zero Postmates spending in more than 50\% of their post-adoption pre-merger months, as these represent users who had effectively disengaged from the platform before the merger.

\section{Methodology}\label{sec:method}

We employ an Age--Period--Cohort (APC) decomposition to identify the causal effect of Uber's acquisition of Postmates (December 2020) on consumer spending allocations. The APC approach identifies the merger effect through a structural break in \emph{period} fixed effects---the time-varying component that remains after absorbing lifecycle (age) and selection (cohort) effects. We implement the cohort-level decomposition similar to the one used in \citet{oblander2023frontiers}.

\subsection{Cohort-Level APC Specification}\label{subsec:om}

We aggregate it to the cohort $\times$ month level, where cohorts are defined by the month of a user's first Postmates order. Let $c$ index cohorts, $t$ calendar months, and $a = t - c$ tenure:
\begin{equation}\label{eq:om_apc}
    \bar{Y}_{cat} = \alpha_{a} + \phi_{c} + \gamma_{t} + \bar{\mathbf{X}}_{cat}'\boldsymbol{\beta} + \varepsilon_{cat}
\end{equation}
where $\bar{Y}_{cat}$ is the cohort-month average outcome, $\alpha_{a}$ are age fixed effects, $\phi_{c}$ are cohort fixed effects, $\gamma_{t}$ are period fixed effects, and $\bar{\mathbf{X}}_{cat}$ are cohort-month averaged controls. Observations are weighted by cohort size $N_{c}$.

The control vector $\bar{\mathbf{X}}_{cat}$ includes three blocks: (i)~\emph{covid controls} (MSA $\times$ time-varying)---standardized total cases, squared cases, workplace closure mandates (partial and full), gathering restrictions (100- and 10-person thresholds), and stay-at-home orders; (ii)~\emph{user $\times$ time controls}---an indicator for active transaction syncing and indicators for concurrent multi-platform usage on UberEats, DoorDash, and GrubHub; and (iii)~\emph{restaurant supply controls}---counts of new and overlapping restaurants on Postmates and competing platforms. All controls are averaged to the cohort-month level; missing values are imputed as zero.

\paragraph{Detrending with seasonality.}  We detrend using a model that accounts for seasonality via a B-spline:
\begin{equation}\label{eq:om_detrend}
    \hat{\gamma}_{t} = a + b \cdot t + \text{bs}(\text{month-of-year}(t), 4) + r_{t}
\end{equation}
where $\text{bs}(\cdot, 4)$ denotes a B-spline with 4 internal knots on the month-of-year, fit on pre-merger data only. in appendix \ref{sec:robust} we look at sensitivity to detrending by varying the number of knots. 

\paragraph{ARIMA counterfactual.} We model the pre-merger residuals as an AR($p$) process:
\begin{equation}\label{eq:om_arima}
    \hat{r}_{t} = \sum_{j=1}^{p} \phi_{j} \hat{r}_{t-j} + \eta_{t}, \quad t < \text{Dec 2020}
\end{equation}
The lag order $p$ is selected by minimizing AIC over $p \in \{0, 1, \ldots, 6\}$. The fitted AR model forecasts the counterfactual post-merger trajectory $\hat{r}^{cf}_{t}$, and the merger lift at each post-merger month is:
\begin{equation}\label{eq:om_lift}
    \text{Lift}_{t} = \hat{r}_{t} - \hat{r}^{cf}_{t}
\end{equation}
In appendix \ref{sec:robust} we provide more details on sensitivity to AR search range and ACF/PACF diagnosis with changes to covariates used. 

\subsection{APC Identification}\label{subsec:apc_id}

The fundamental APC identification challenge is the exact linear dependence $a = t - c$. The separate age, cohort, and period fixed effects are not individually point-identified, but the \emph{detrended} period effects---and hence the structural break at the merger---are identified under the assumption that the pre-merger trend would have continued absent the merger.

\subsection{Validating the Identifying Assumption}\label{subsec:validation}

Following \cite{blanchard2025game}, we validate the APC identifying assumption by examining whether cohorts that experienced the merger at different tenures followed parallel age paths before the merger. If the age effects are stable across cohorts, any post-merger divergence can be attributed to the period (merger) effect rather than confounding cohort--age interactions.

\begin{figure}[H]
    \centering
    \includegraphics[width=\textwidth]{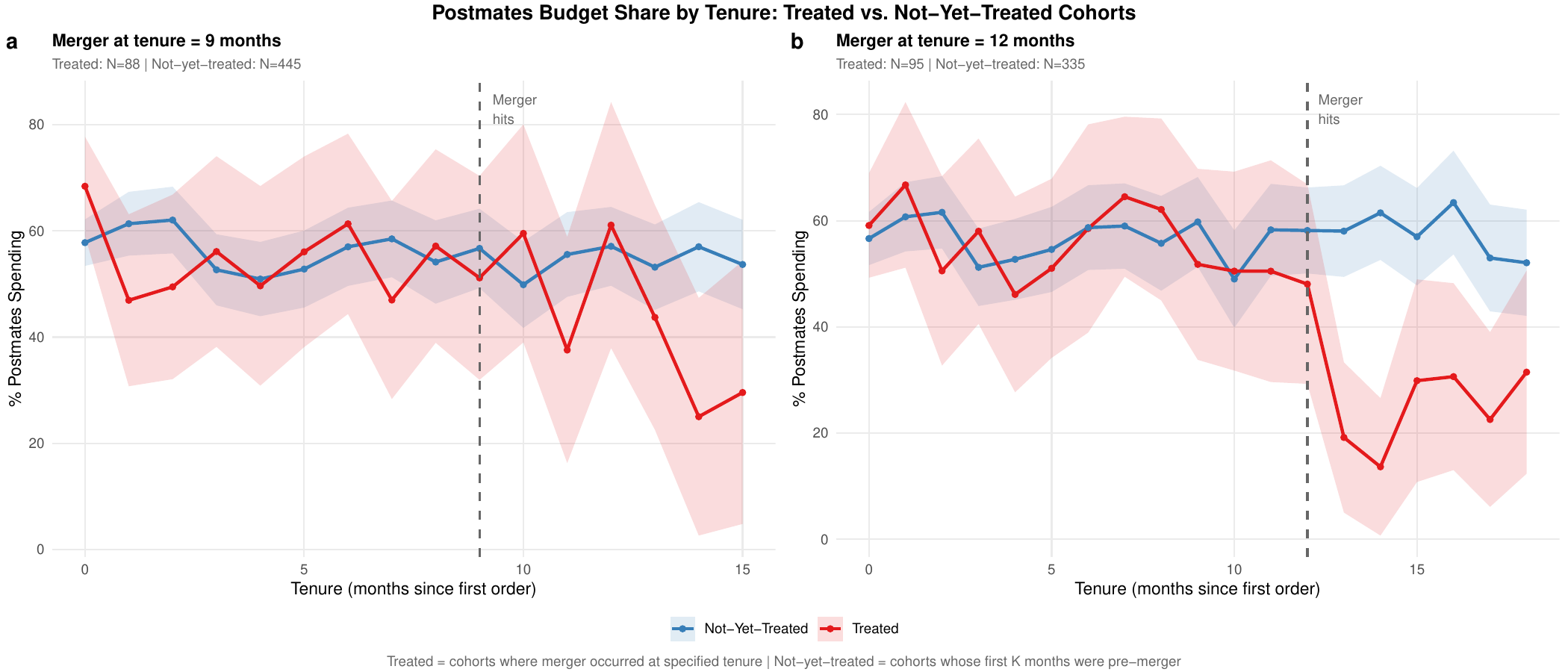}
    \caption{}
        \label{fig:bp_fig3}
\end{figure}

Figure~\ref{fig:bp_fig3} compares the raw Postmates budget share by tenure for ``treated'' cohorts (those who experienced the merger at a given tenure) versus ``not-yet-treated'' cohorts (those who had already passed that tenure before the merger). The two panels show merger impacts at 9 and 12 months of tenure. Pre-merger parallel paths support the identifying assumption.

Postmates budget share by tenure: treated vs.\ not-yet-treated cohorts. Treated cohorts experienced the merger at the indicated tenure; not-yet-treated cohorts had already passed that tenure pre-merger. Parallel pre-merger paths validate the APC identifying assumption. Additional validation using pre-merger outcomes by cohort and tenure is presented in Appendix~\ref{app:validation}.

\section{Results}\label{sec:results}

\subsection{Event Study}\label{subsec:main_results}

Table~\ref{tab:om_shifts} reports the estimated post-merger shifts for all dependent variables. Each column corresponds to a different outcome; the coefficient represents the average gap between actual and counterfactual period effects over the post-merger window, with standard errors in parentheses. The bottom panel indicates the fixed effects and controls included.  We see that postmates users, decreased their budget, on average by $40\%$, which led to gain in budget allocated to the acquired platform i.e,\ UberEats ($19.8\%$) but also led to increase in budget percentage spent on competing firms Doordash ($7.5\%$) and Grubhub ($10.5\%$).

\input{tex/APC_table2_OM_shifts.tex}

\begin{figure}[H]
    \centering
    \includegraphics[width=\textwidth]{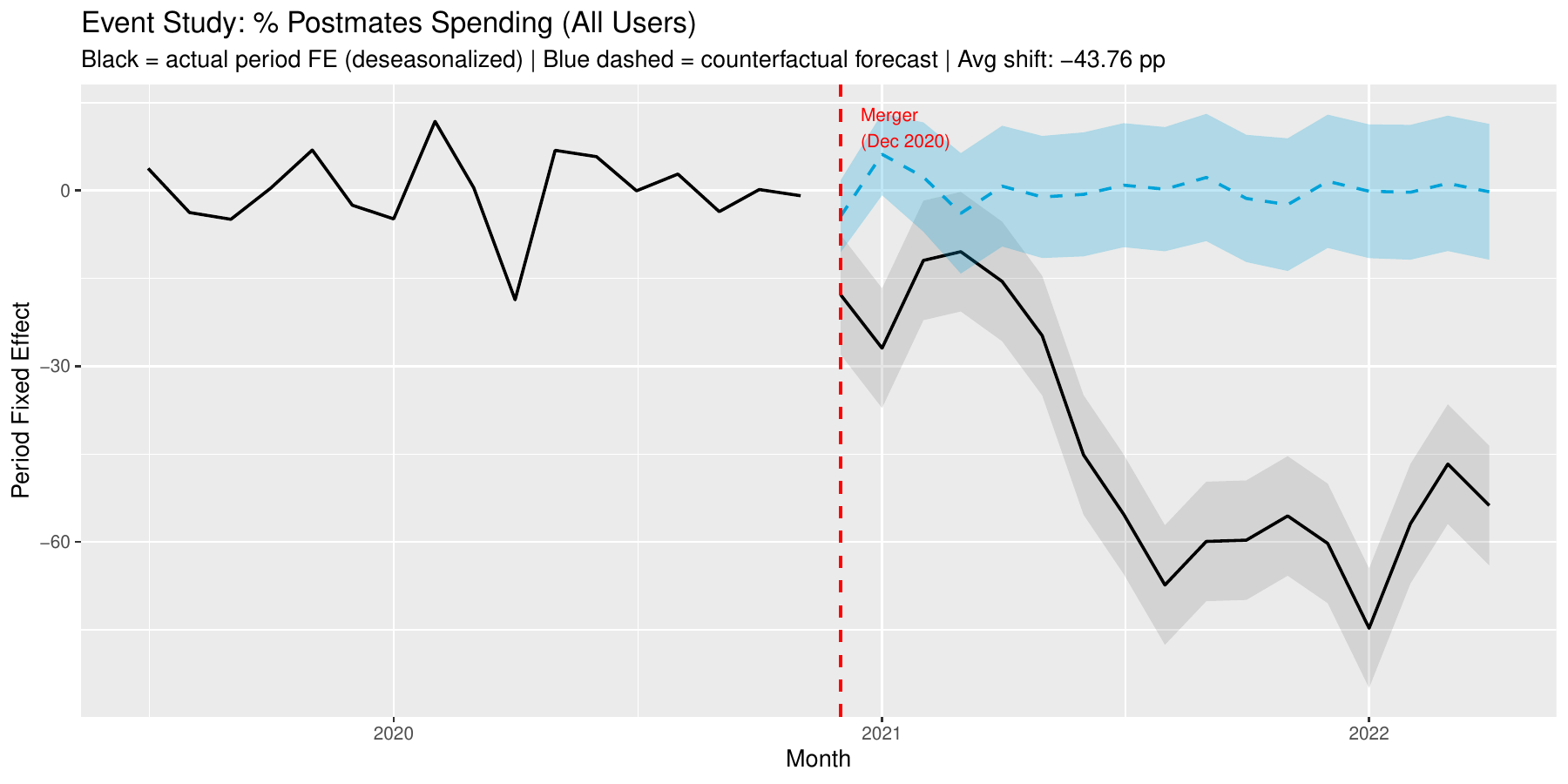}
    \caption{Event study: actual vs.\ counterfactual detrended period fixed effects for Postmates budget share. The counterfactual is generated from an AR model fit on pre-merger residuals. Shaded bands show 95\% confidence intervals. The gap post-December 2020 estimates the merger effect.}
    \label{fig:om_main}
\end{figure}

Figure~\ref{fig:om_main} presents the event study for Postmates budget share. The solid line shows the actual detrended period fixed effects; the dashed line shows the AR-based counterfactual forecast. The shaded regions represent 95\% confidence intervals. The gap between the actual and counterfactual series after December 2020 represents the estimated merger effect.

\begin{figure}[H]
    \centering
    \includegraphics[width=\textwidth]{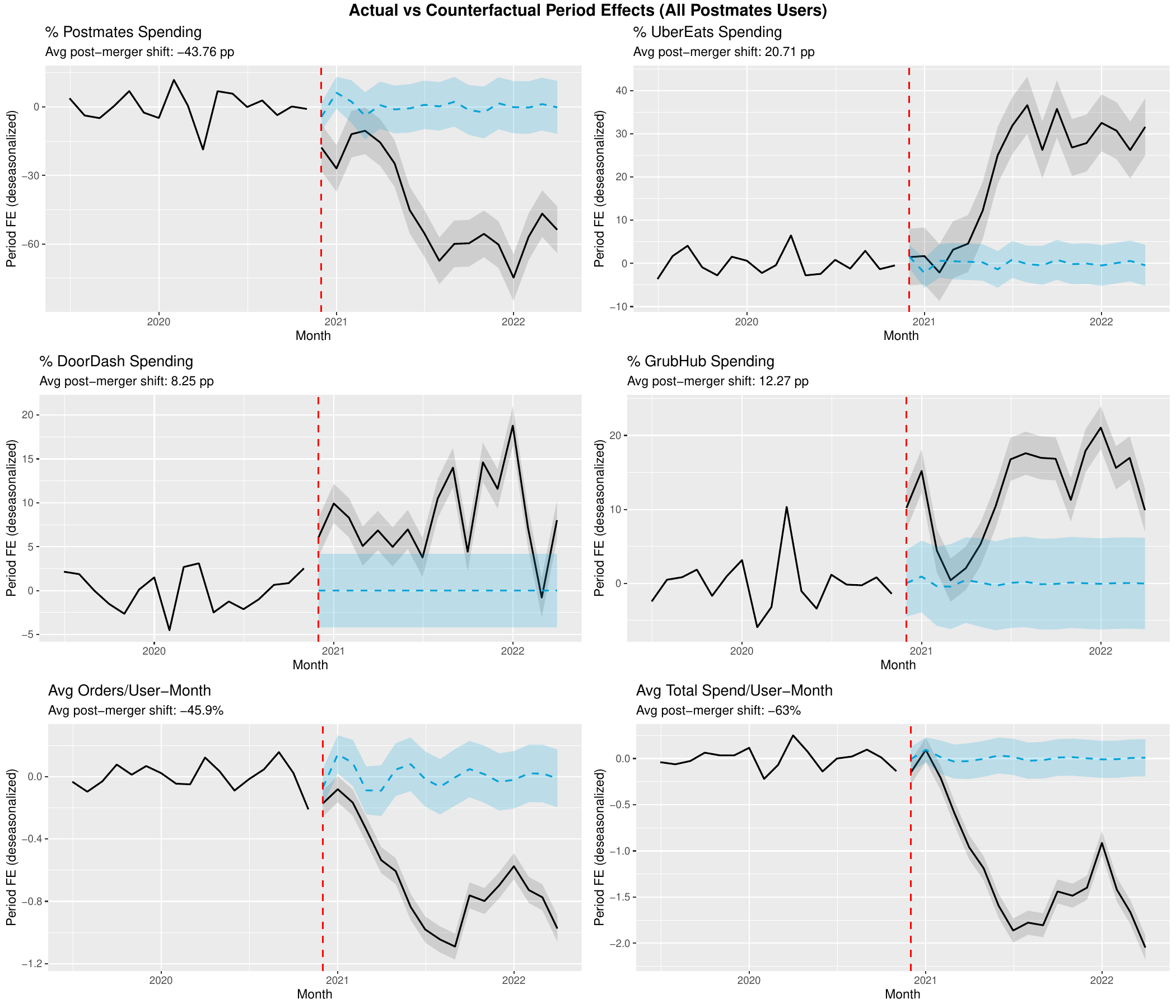}
    \caption{Event studies across all dependent variables. Each panel shows actual vs.\ counterfactual detrended period FEs.}
    \label{fig:om_all_dvs}
\end{figure}

Figure~\ref{fig:om_all_dvs} extends the analysis to all dependent variables: budget shares for each platform (Postmates, UberEats, DoorDash, GrubHub) as well as total spending and order counts.

\subsection{Heterogeneous Effects}\label{subsec:het_results}

\begin{figure}[H]
    \centering
    \begin{subfigure}[b]{0.48\textwidth}
        \centering
        \includegraphics[width=\textwidth]{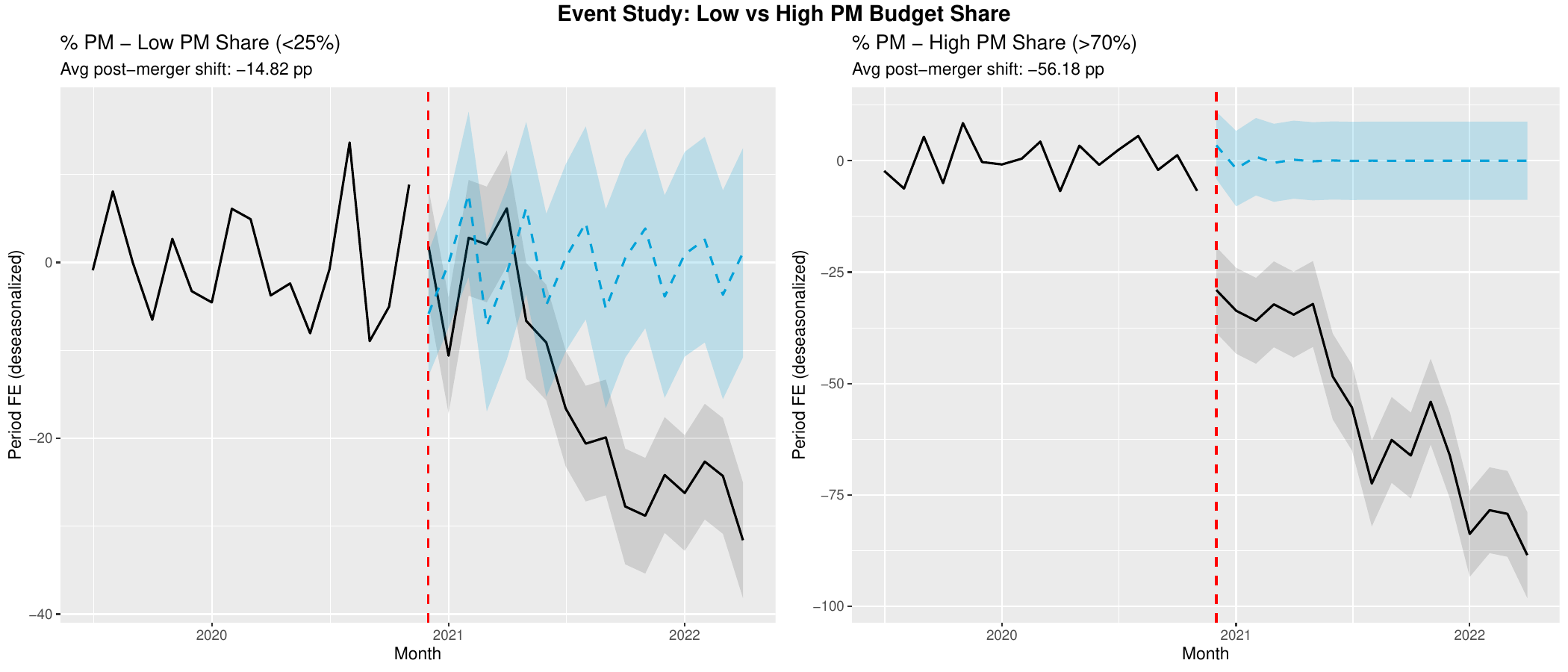}
        \caption{By budget share}
    \end{subfigure}
    \hfill
    \begin{subfigure}[b]{0.48\textwidth}
        \centering
        \includegraphics[width=\textwidth]{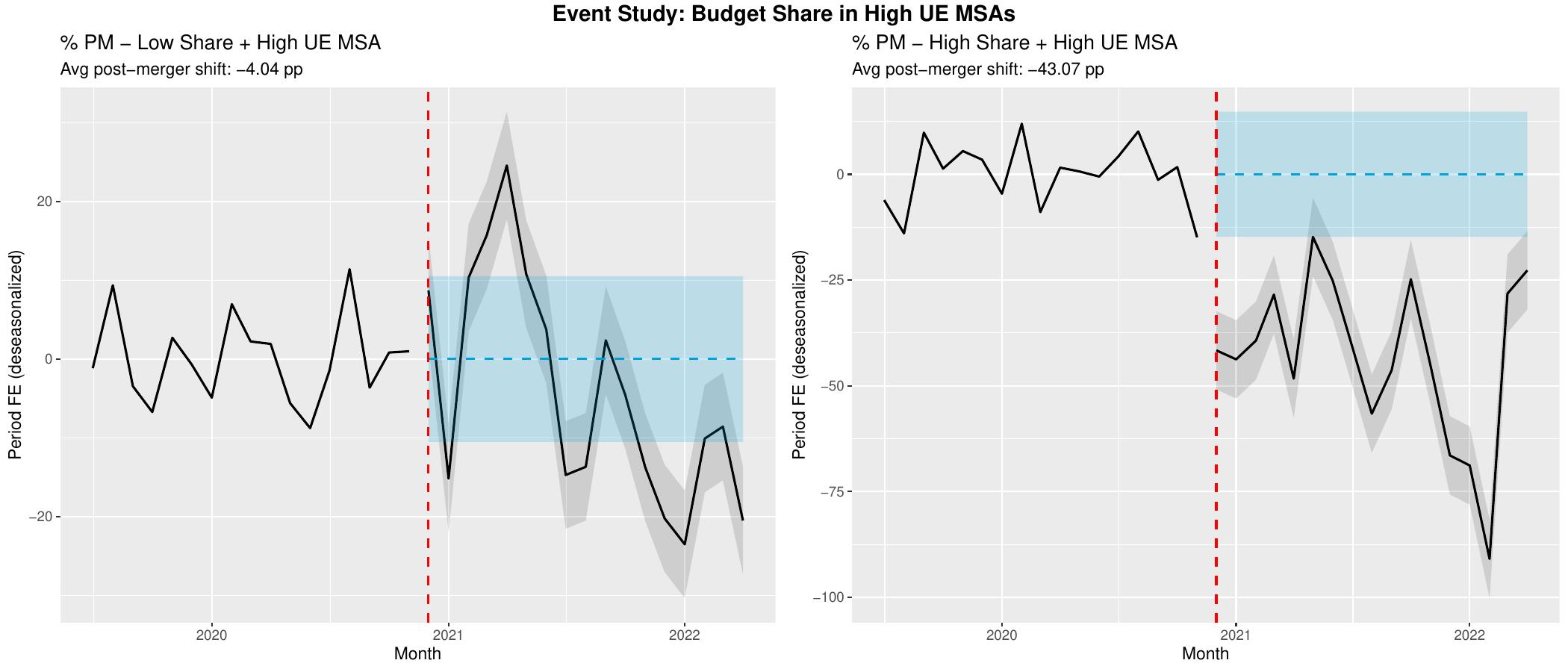}
        \caption{Budget share $\times$ High UE MSA}
    \end{subfigure}
    \caption{Heterogeneous event studies. Each panel shows actual vs.\ counterfactual detrended period fixed effects with 95\% confidence intervals for the indicated subgroup split.}
    \label{fig:het_main}
\end{figure}

We examine heterogeneity along two dimensions: pre-merger Postmates budget share and the interaction of budget share with local UberEats market concentration. Figure~\ref{fig:het_main} presents the event studies for these subgroups. Additional heterogeneity analyses by multi-homing status, MSA market structure, and their interactions are presented in Appendix~\ref{app:het_additional}.

Table~\ref{tab:het_summary} summarizes the post-merger shift estimates across all subgroups.

\input{tex/APC_table4_het_summary.tex}

\subsection{Comparison with Difference-in-Differences}\label{subsec:did_comparison}

In this section, we complement the main results by using a difference in difference approach, which identifies the merger effect through cross-market variation in competitive exposure: MSAs are assigned to the treatment group if their UberEats market share exceeds the sample median:
\begin{equation}\label{eq:treatment}
    \text{Treated}_{m} = \mathbf{1}\left[\text{UE\_Share}_{m} > \text{Median}(\text{UE\_Share})\right]
\end{equation}
The estimating equation is a two-way fixed effects difference-in-differences model:
\begin{equation}\label{eq:did_main}
    y_{imt} = \alpha + \beta \cdot (\text{Treated}_{m} \times \text{Post}_{t}) + \mathbf{X}'_{imt}\boldsymbol{\gamma} + \mu_{m} + \lambda_{t} + \varepsilon_{imt}
\end{equation}
where $y_{imt}$ is the percentage of total food delivery spending allocated to Postmates by consumer $i$ in MSA $m$ at month $t$; $\text{Treated}_{m}$ is the binary treatment indicator defined in Equation~\eqref{eq:treatment}; $\text{Post}_{t} = \mathbf{1}[t \geq \text{Dec 2020}]$ indicates the post-merger period; $\mu_{m}$ and $\lambda_{t}$ are MSA and year-month fixed effects, respectively; and $\mathbf{X}_{imt}$ includes controls added cumulatively across eight specifications: sync status, Postmates tenure, COVID-19 restrictions, multi-platform indicators, budget share categories, and restaurant supply variables. Standard errors are clustered at the MSA level.  Note that here coefficient $\beta$ captures the average treatment effect on the treated---the differential change in Postmates spending share for consumers in high-UberEats-share MSAs relative to those in low-UberEats-share MSAs, after the merger.

\begin{figure}[H]
    \centering
    \includegraphics[width=0.95\textwidth]{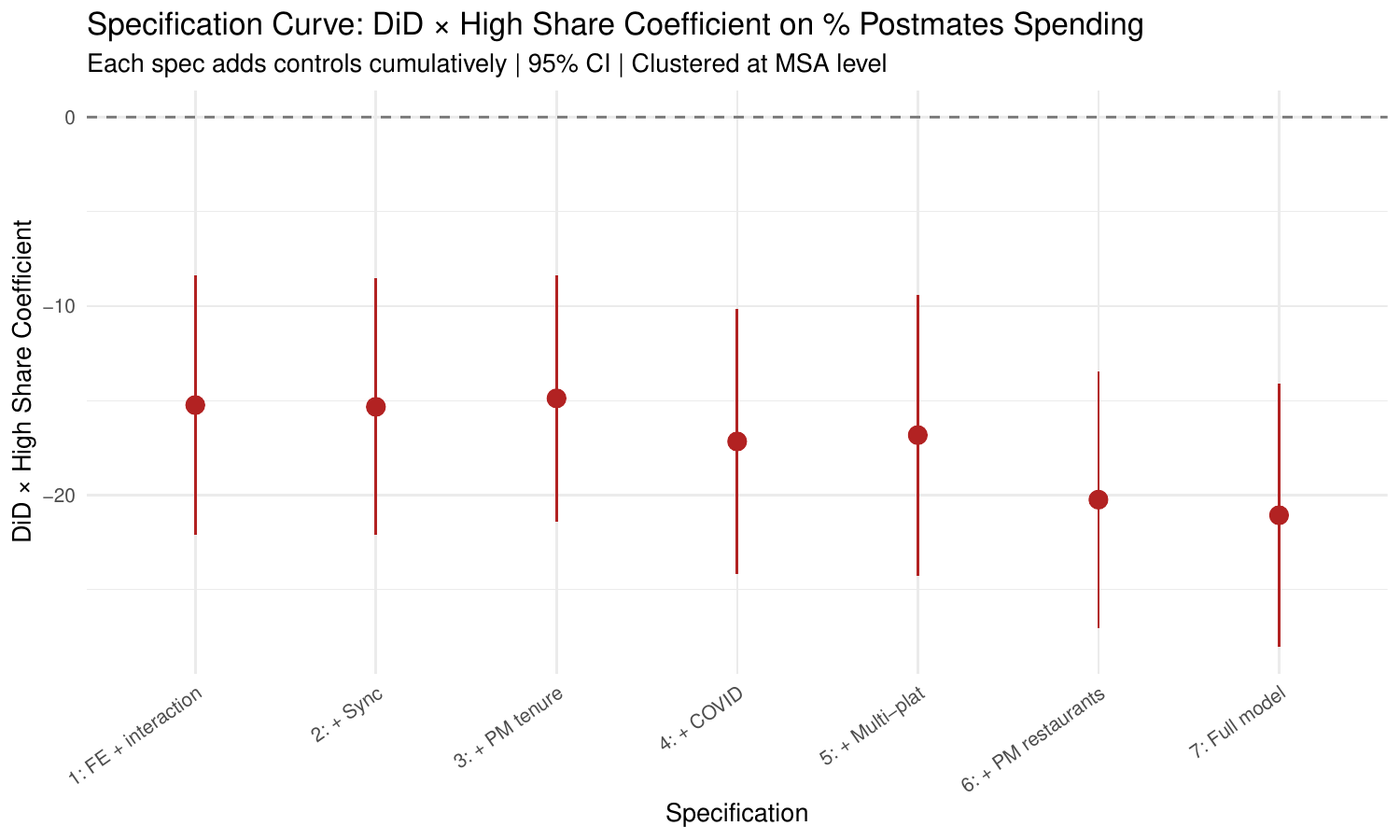}
    \caption{Specification Curve: DiD $\times$ High Share Coefficient. The coefficient on the interaction between the DiD term and an indicator for consumers with $>$70\% pre-merger Postmates spending share.}
    \label{fig:spec_curve_high_share}
\end{figure}

\paragraph{Heterogeneous DiD.} Similar to our APC estimate we also estimates heterogeneous  treatment effect for high pre-merger Postmates dependence as captured by higher than 70\% of the online food delivery budget allocated to Postmates, i.e.\ ($s_{i,\text{PM}} > 70\%$):
\begin{equation}\label{eq:did_het_om}
    y_{imt} = \alpha + \beta_{1} \cdot \text{DiD}_{mt} + \beta_{2} \cdot \text{DiD}_{mt} \times \text{HighShare}_{i} + \delta_{1} \cdot \text{LowShare}_{i} + \delta_{2} \cdot \text{HighShare}_{i} + \mathbf{X}_{imt}'\boldsymbol{\gamma} + \mu_{m} + \lambda_{t} + \varepsilon_{imt}
\end{equation}
where $\text{HighShare}_{i} = \mathbf{1}[s_{i,\text{PM}} > 70\%]$ and $\text{LowShare}_{i} = \mathbf{1}[s_{i,\text{PM}} < 25\%]$. The specification curve for $\hat{\beta}_{2}$ (Figure~\ref{fig:spec_curve_high_share}) captures the \emph{additional} effect of the merger on high-dependence consumers in treated MSAs relative to low/medium-share consumers.

\paragraph{DiD results} 

We present results from a specification curve analysis that progressively adds control variables to the baseline two-way fixed effects model. This approach demonstrates the stability of the estimates across different modeling choices.

\begin{figure}[H]
    \centering
    \includegraphics[width=0.95\textwidth]{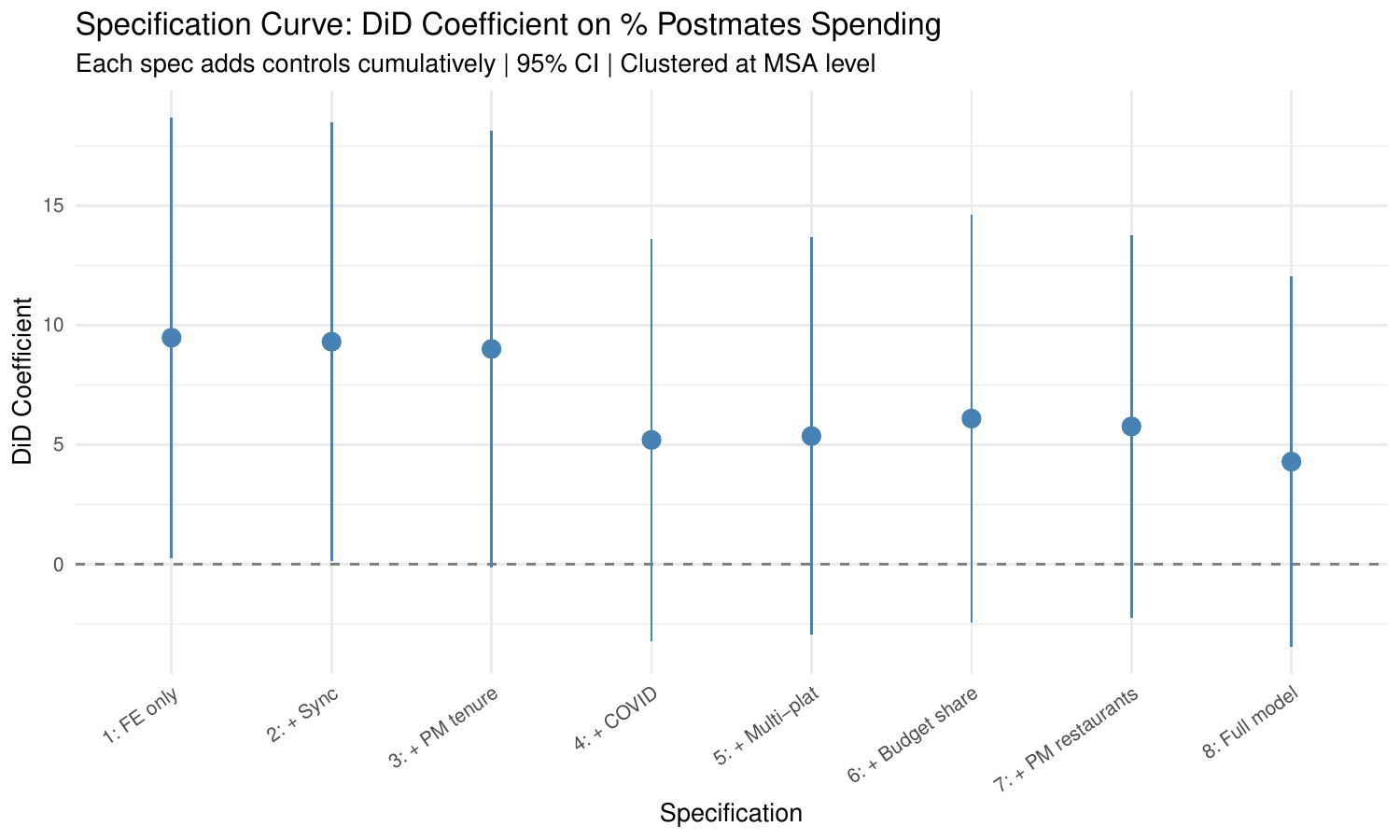}
    \caption{Specification Curve: DiD Coefficient on Percentage Postmates Spending. Each specification cumulatively adds controls. Error bars indicate 95\% confidence intervals with standard errors clustered at the MSA level.}
    \label{fig:spec_curve_did}
\end{figure}

The specification curve in Figure~\ref{fig:spec_curve_did} plots the DiD coefficient $\hat{\beta}$ from Equation~\eqref{eq:did_main} across eight specifications that progressively add controls: (1)~MSA and time fixed effects only; (2)~sync status; (3)~Postmates tenure; (4)~COVID-19 controls; (5)~multi-platform indicators; (6)~budget share categories; (7)~Postmates restaurant controls; and (8)~the full model including other-platform restaurant controls.

{Heterogeneous Effects: High Postmates Share Interaction}\label{subsec:did_het_results}

Figure~\ref{fig:spec_curve_high_share} presents the specification curve for the interaction coefficient $\hat{\beta}_{2}$ from Equation~\eqref{eq:did_het_om}, capturing the differential effect on high-dependence Postmates consumers.\

\paragraph{Comparison of approaches.}
The DiD and the cohort-level APC decomposition identify the merger effect through fundamentally different sources of variation. The DiD exploits \emph{cross-sectional} variation---comparing high- vs.\ low-UberEats markets before and after December 2020. By construction, the DiD coefficient $\hat{\beta}$ captures only the \emph{differential} effect in high-UberEats markets relative to low-UberEats markets. If the merger also affected spending in low-UberEats markets (e.g., through platform-wide integration, branding changes, or supply-side consolidation), the DiD understates the total effect. The cohort-level APC decomposition (Section~\ref{sec:method}), in contrast, exploits \emph{time-series} variation, identifying the merger through a structural break in period fixed effects after absorbing lifecycle (age) and selection (cohort) effects. It therefore captures the \emph{overall} merger effect on all Postmates users regardless of local market structure.

\paragraph{Heterogeneous effects comparison.} Both approaches find that the merger's impact varies with users' pre-merger Postmates dependence, but they measure different margins. The DiD interaction coefficient $\hat{\beta}_{2}$ measures the \emph{additional} effect of high Postmates spending share \emph{within treated markets only}---it asks whether high-share users in high-UberEats MSAs experienced a larger decline than low-share users in those same MSAs. The APC heterogeneity results (Table~\ref{tab:het_summary} and Figure~\ref{fig:het_main}) instead measure the post-merger shift in \emph{levels} for each subgroup across the full sample. The APC results by budget share $\times$ high-UE MSA (Figure~\ref{fig:het_main}b) show that the merger effect is most pronounced among high-share users in markets where Uber had the strongest competitive presence---precisely the subgroup where the DiD treatment bite is largest. This alignment across the two identification strategies reinforces the finding that consumers most dependent on Postmates experienced the largest budget reallocation following the merger. However, the APC approach complements the DiD by additionally revealing the \emph{magnitude} of the overall merger-induced shift, while the DiD provides a useful lower bound on the treatment effect in the most directly affected markets.

\section{Conclusion}\label{sec:conclusion}

This paper examines the consumer-side consequences of Uber's acquisition of Postmates using an Age--Period--Cohort decomposition that isolates the merger's causal effect from lifecycle and selection dynamics. Our results yield three main findings with implications for understanding horizontal acquisitions in multi-sided platform markets.

First, the merger caused a substantial decline in Postmates budget share ($-40.33$ percentage points), confirming that the acquisition disrupted existing consumer relationships with the target platform. However, this spending reduction did not translate into a corresponding increase in UberEats usage. While UberEats gained $+19.79$ pp, DoorDash ($+7.50$ pp) and Grubhub ($+10.49$ pp) collectively absorbed a significant share of the reallocated spending. This finding suggests that in markets with readily available substitutes, horizontal acquisitions may inadvertently benefit rival platforms by disrupting consumer habits without fully redirecting demand toward the acquirer.

Second, our heterogeneity analysis reveals that the merger's impact was concentrated among high-dependence Postmates users, particularly those in markets with strong UberEats presence ($-88.18$ pp for high-share single-homers in high-UE MSAs). In contrast, multi-homing consumers with diversified platform usage proved more resilient to the merger shock ($-35.14$ pp). This pattern implies that consumers with established relationships across multiple platforms are less sensitive to acquisition-induced disruptions, which has implications for how firms should sequence post-merger integration strategies.

Third, our comparison of the APC approach with a standard Difference-in-Differences design demonstrates that the two methods are complementary. The DiD captures the differential effect across markets with varying competitive exposure, while the APC recovers the total merger-induced shift in consumer behavior. The fact that the APC estimates exceed the DiD estimates suggests that a substantial portion of the merger's effect operates through platform-wide channels---such as branding changes, app integration, and supply-side consolidation---that are common across all markets and therefore missed by cross-sectional comparisons.

Our findings carry implications for both merger policy and platform strategy. For antitrust authorities, the evidence that displaced demand flows to competitors rather than to the acquirer suggests that horizontal platform mergers may be less anticompetitive than traditional market concentration metrics would imply. However, the welfare implications depend on whether rival platforms offer comparable quality and variety. For platform operators considering horizontal acquisitions, our results highlight the risk that integration-induced disruptions can erode the acquired customer base before the intended demand consolidation materializes. The stickiness of multi-homing consumers suggests that more diversified users are more "sticky" in their demand and might be less likely to switch platforms.

\clearpage
\bibliographystyle{aer}
\bibliography{bib}
\newpage 
\appendix
\section{Appendix}
\section{Additional Validation}\label{app:validation}

\begin{figure}[H]
    \centering
    \includegraphics[width=\textwidth]{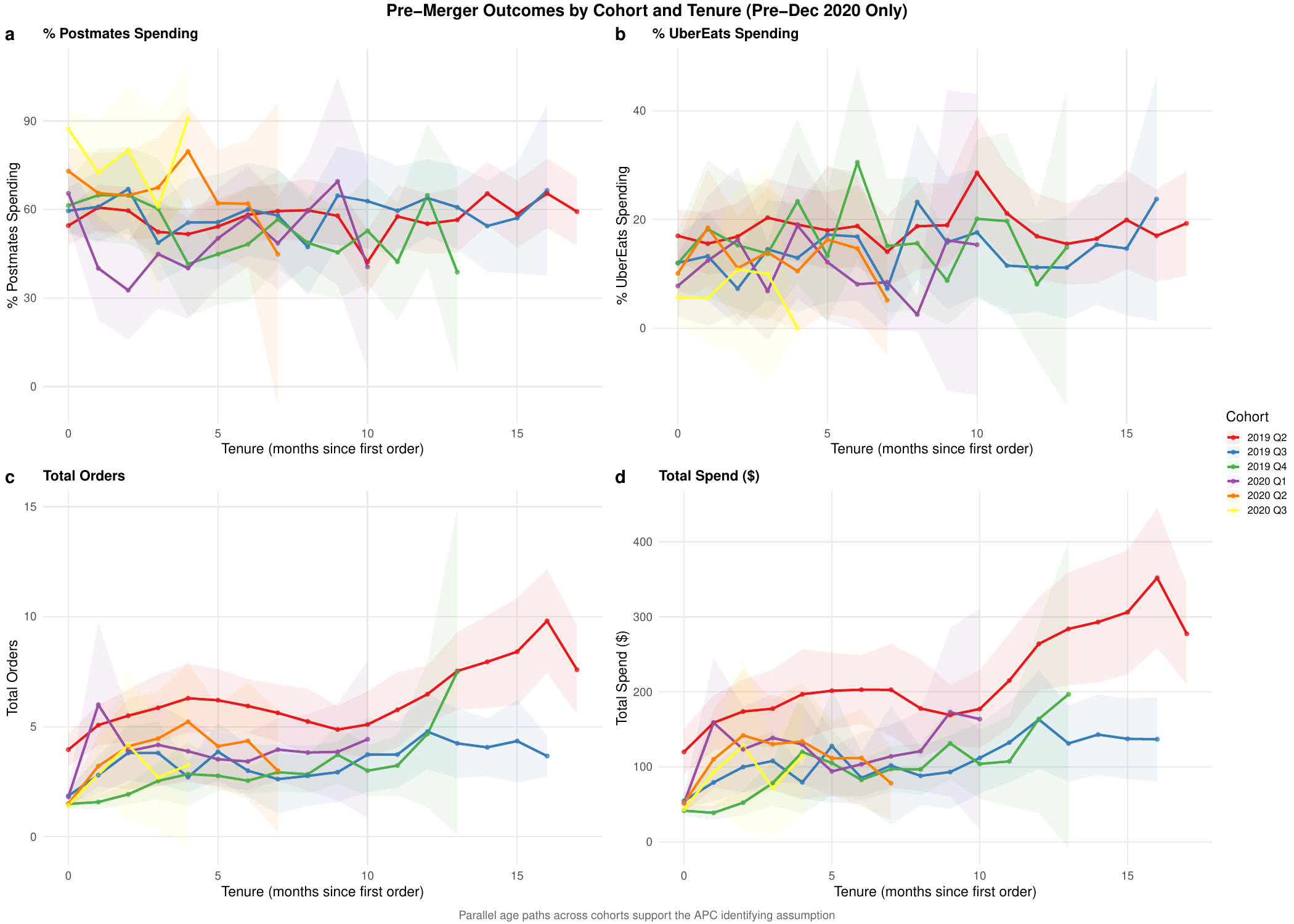}
    \caption{Pre-merger budget share outcomes by cohort and tenure (pre-Dec 2020 only). Parallel age paths across quarterly cohorts support the APC identifying assumption.}
    \label{fig:bp_fig4}
\end{figure}

Figure~\ref{fig:bp_fig4} plots pre-merger outcomes by cohort and tenure using only pre-December 2020 data. Cohorts are grouped by quarter of first order. Nearly identical age paths across cohorts confirm that lifecycle patterns are stable---a necessary condition for the APC decomposition to correctly attribute post-merger shifts to the period effect.

\subsection{Specification Curve}

\begin{figure}[H]
    \centering
    \includegraphics[width=\textwidth]{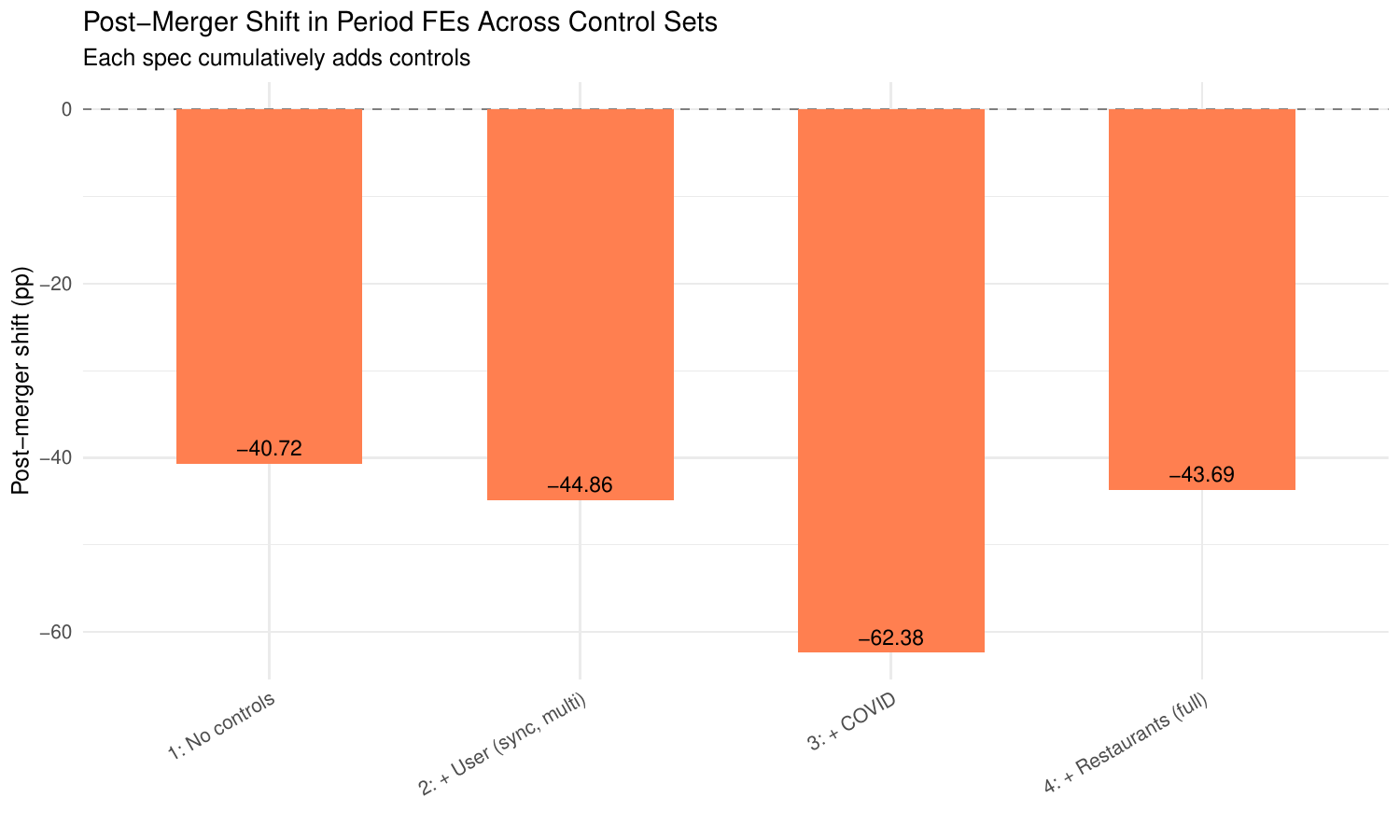}
    \caption{Specification curve: Post-merger shift across control sets.}
    \label{fig:om_spec}
\end{figure}

Figure~\ref{fig:om_spec} and Table~\ref{tab:om_spec_curve} present the specification curve.

\input{tex/APC_table6_OM_spec_curve_controls.tex}

\section{Robustness Checks}\label{sec:robust}

\subsection{Sensitivity to AR Search Range}\label{subsec:robust_ar}

The counterfactual depends on the AR($p$) model selected by AIC. We assess sensitivity to the search range by comparing AR orders selected from $\{0, \ldots, 3\}$, $\{0, \ldots, 6\}$, and $\{0, \ldots, 9\}$.

\begin{figure}[H]
    \centering
    \includegraphics[width=\textwidth]{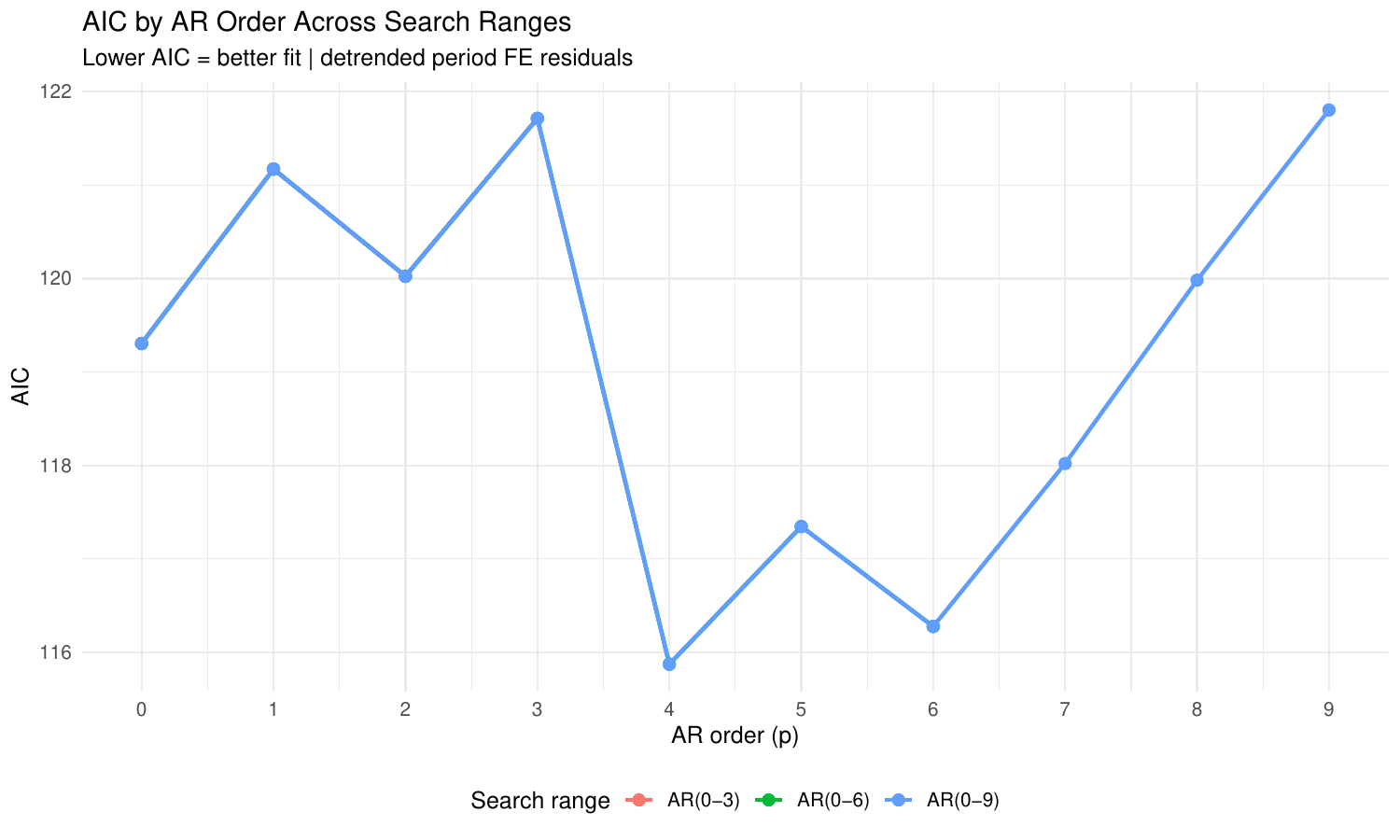}
    \caption{AIC by AR order across search ranges. Lower AIC indicates better fit.}
    \label{fig:ar_aic}
\end{figure}

Figure~\ref{fig:ar_aic} plots AIC as a function of AR order across all three ranges. Table~\ref{tab:robust_ar} reports the selected model and resulting shift estimate.

\input{tex/APC_robust_table1_AR_sensitivity.tex}

\begin{figure}[H]
    \centering
    \includegraphics[width=0.85\textwidth]{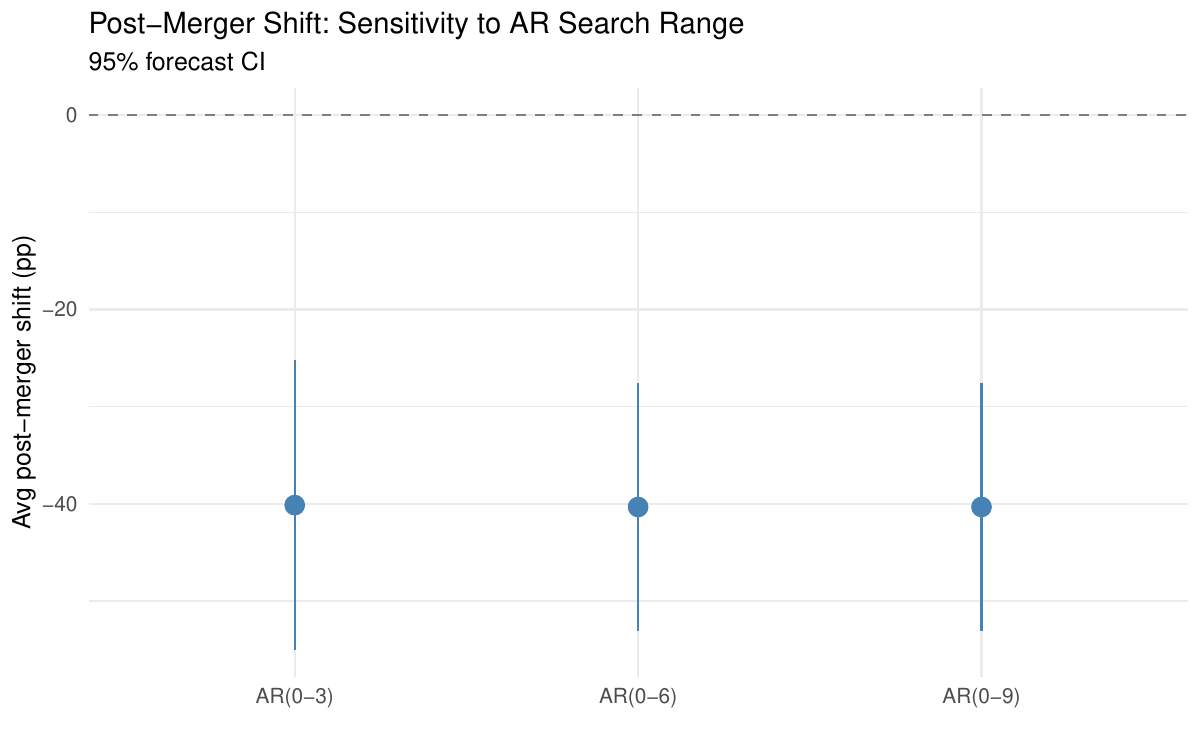}
    \caption{Post-merger shift: sensitivity to AR search range. Error bars show 95\% forecast CIs.}
    \label{fig:ar_shift}
\end{figure}

\begin{figure}[H]
    \centering
    \includegraphics[width=\textwidth]{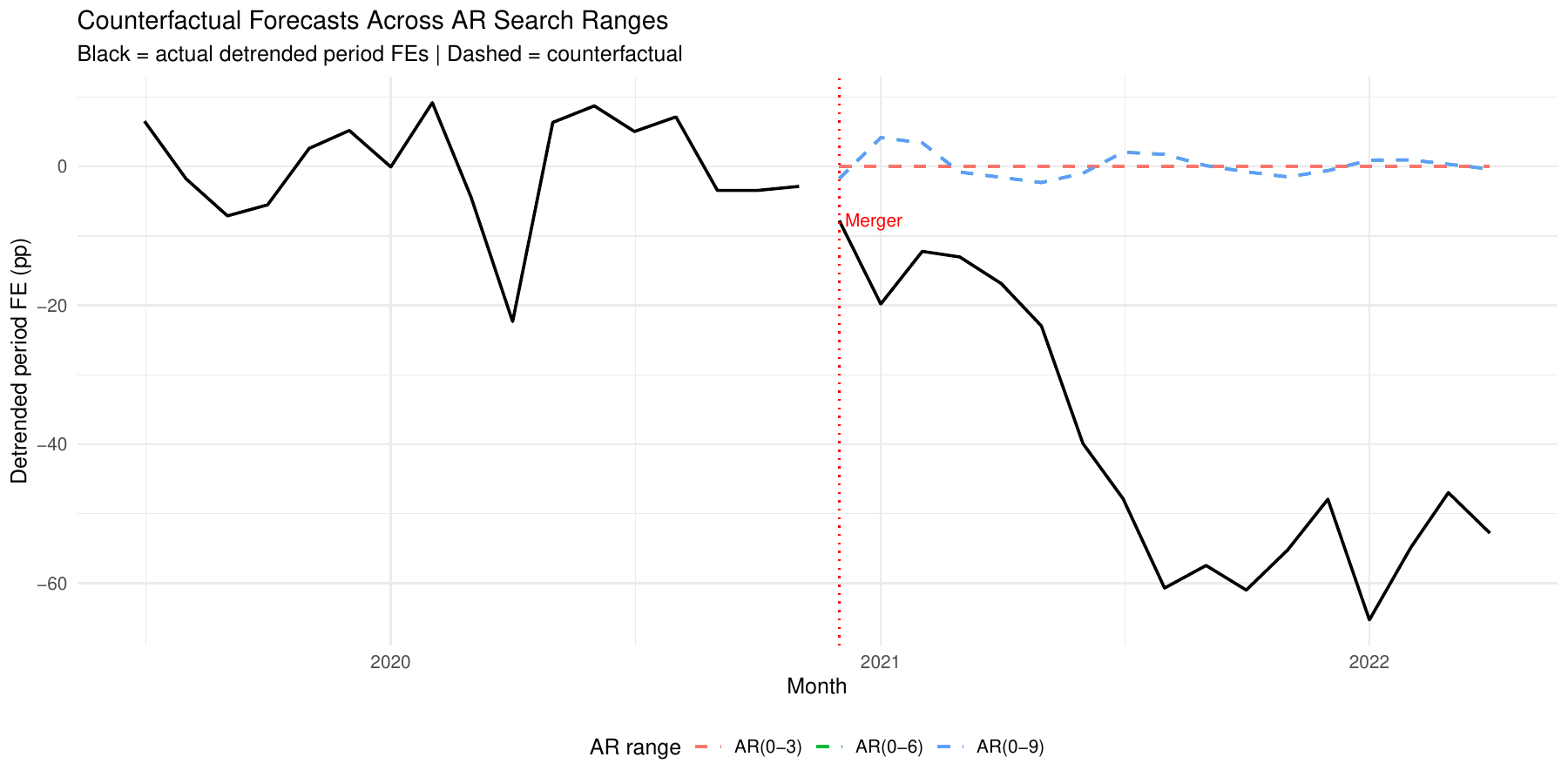}
    \caption{Counterfactual forecasts across AR search ranges overlaid on actual detrended period FEs.}
    \label{fig:ar_cf}
\end{figure}

Figure~\ref{fig:ar_shift} compares the post-merger shift estimates, and Figure~\ref{fig:ar_cf} overlays the counterfactual forecasts from each AR range.

\subsection{Sensitivity to Seasonality Detrending}\label{subsec:robust_knots}

The detrending step uses a B-spline on month-of-year with a default of 3 knots. We vary the number of knots from 3 to 6 to assess sensitivity.

\begin{figure}[H]
    \centering
    \includegraphics[width=0.85\textwidth]{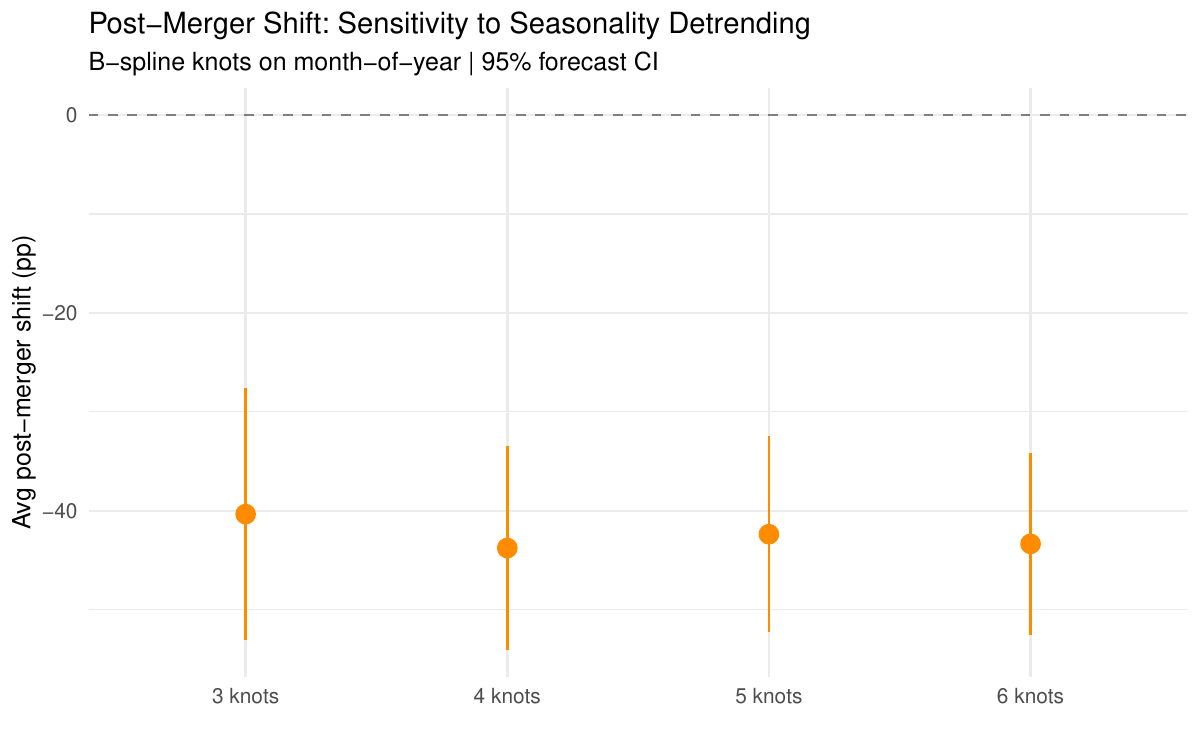}
    \caption{Post-merger shift: sensitivity to B-spline knots in seasonality detrending.}
    \label{fig:knot_shift}
\end{figure}

\begin{figure}[H]
    \centering
    \includegraphics[width=\textwidth]{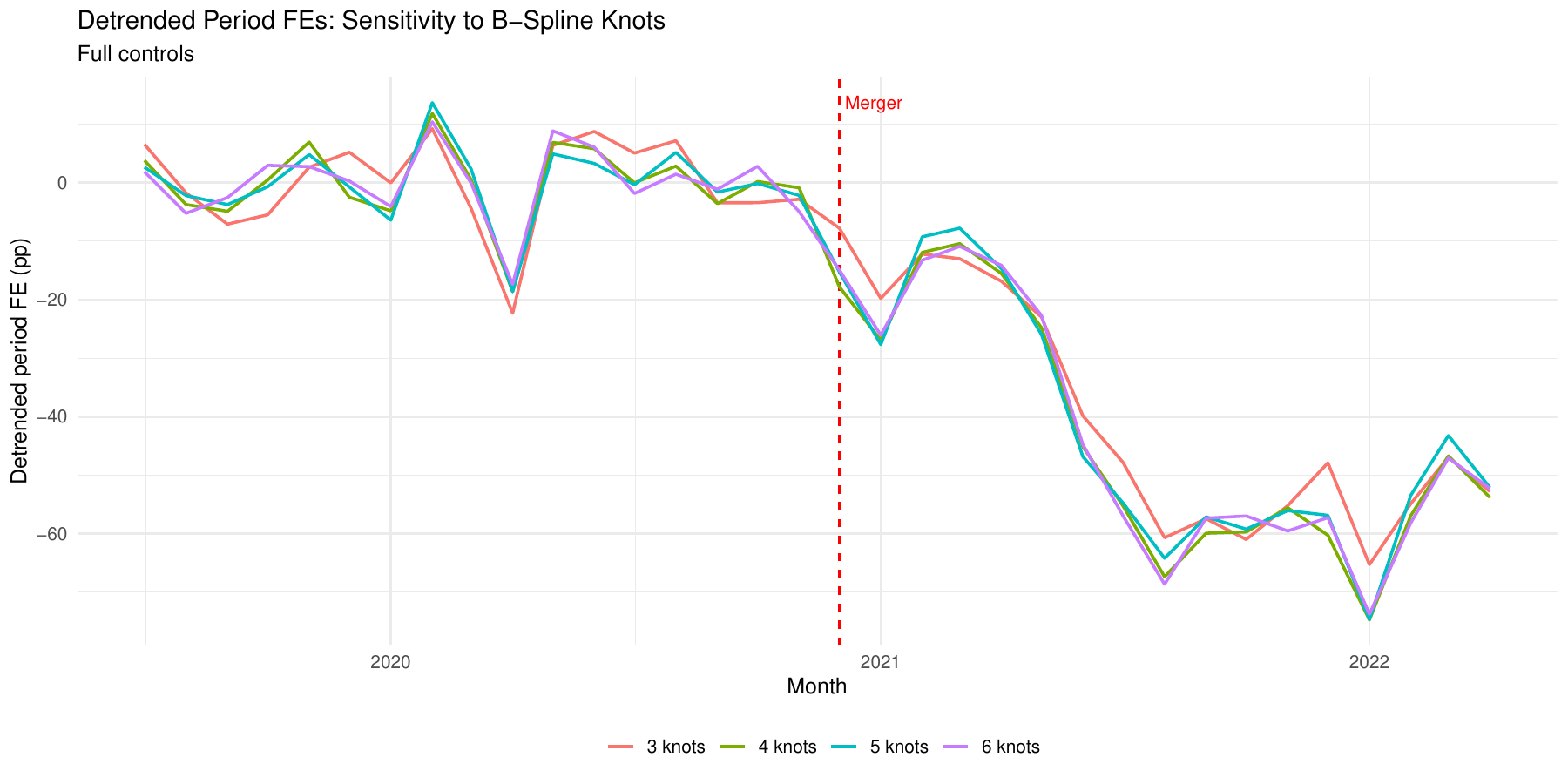}
    \caption{Detrended period FEs across different B-spline knot specifications.}
    \label{fig:knot_overlay}
\end{figure}

Table~\ref{tab:robust_knots} reports results. Figure~\ref{fig:knot_shift} compares shift estimates and Figure~\ref{fig:knot_overlay} overlays the detrended period FEs.

\input{tex/APC_robust_table2_knot_sensitivity.tex}

\subsection{Controls Specification Curve}\label{subsec:robust_controls}

\begin{figure}[h!]
    \centering
    \includegraphics[width=\textwidth]{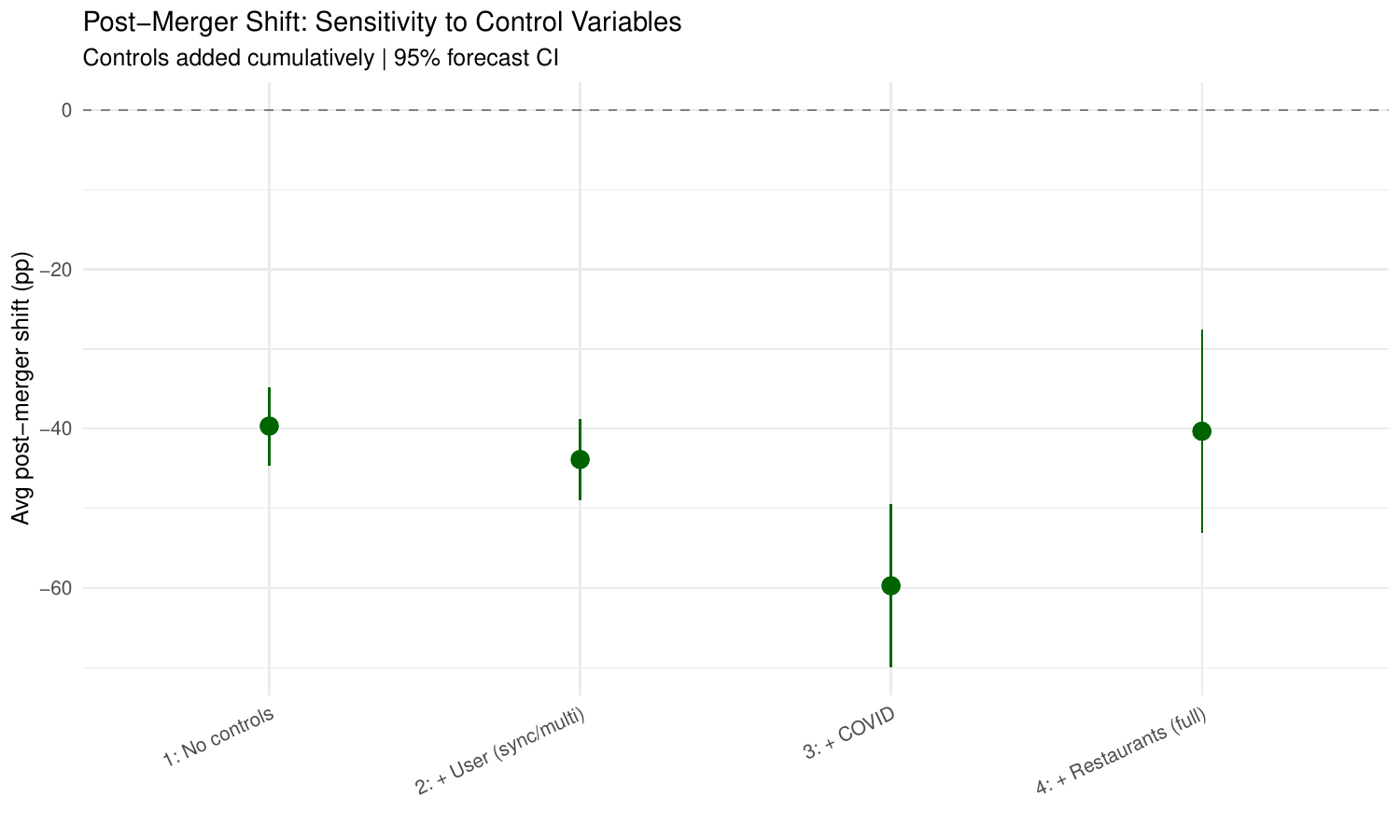}
    \caption{Specification curve: Post-merger shift across cumulative control sets.}
    \label{fig:ctrl_spec}
\end{figure}

Table~\ref{tab:robust_controls} reports the shift as controls are progressively added (no controls $\to$ user $\to$ COVID $\to$ restaurants). Figure~\ref{fig:ctrl_spec} visualizes the specification curve.

\input{tex/APC_robust_table3_controls_spec.tex}

\subsection{ACF and PACF Diagnostics}\label{subsec:robust_acf}

\begin{figure}[H]
    \centering
    \includegraphics[width=0.9\textwidth]{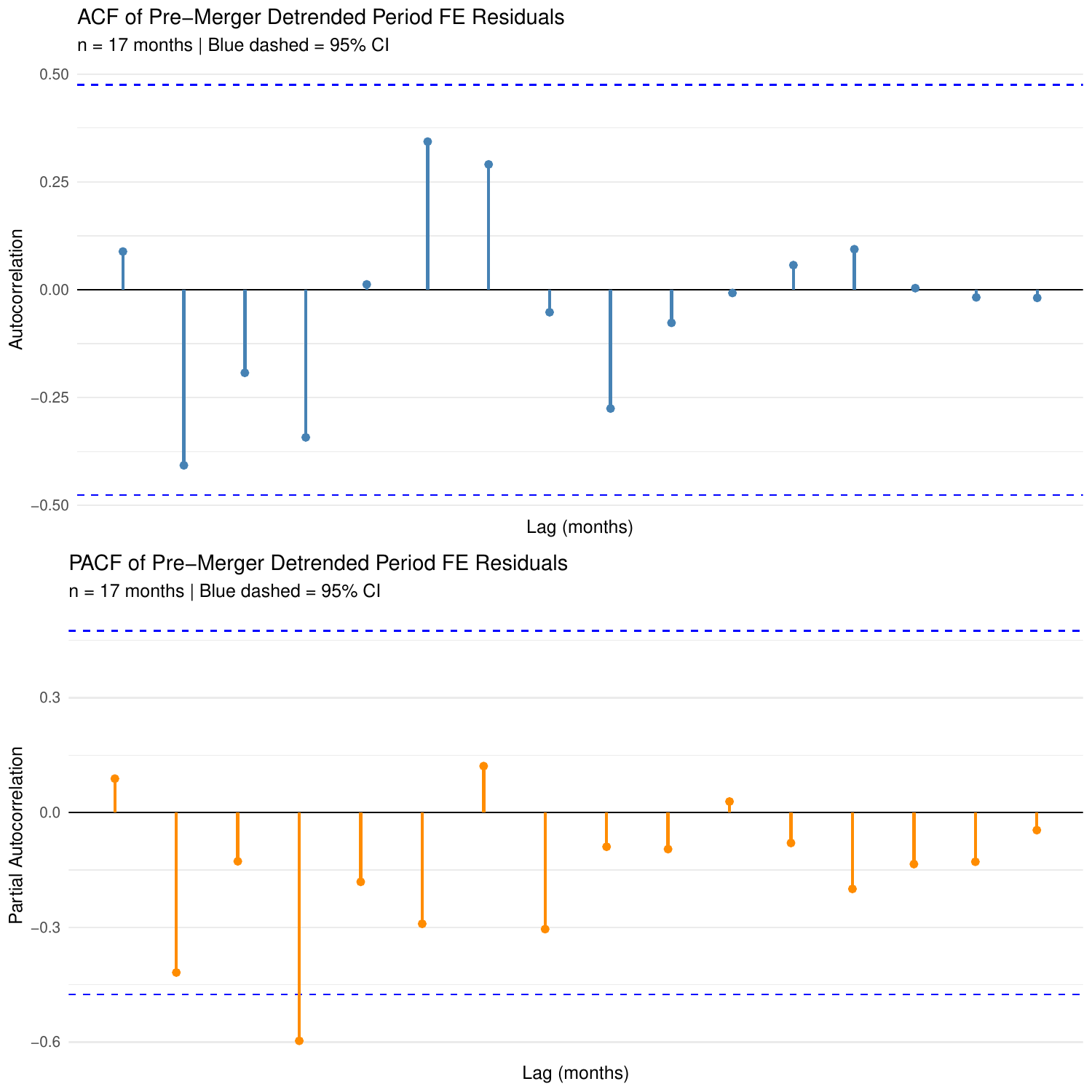}
    \caption{ACF and PACF of pre-merger detrended period FE residuals. Blue dashed lines indicate 95\% confidence bounds.}
    \label{fig:acf_pacf}
\end{figure}

\begin{figure}[H]
    \centering
    \includegraphics[width=0.9\textwidth]{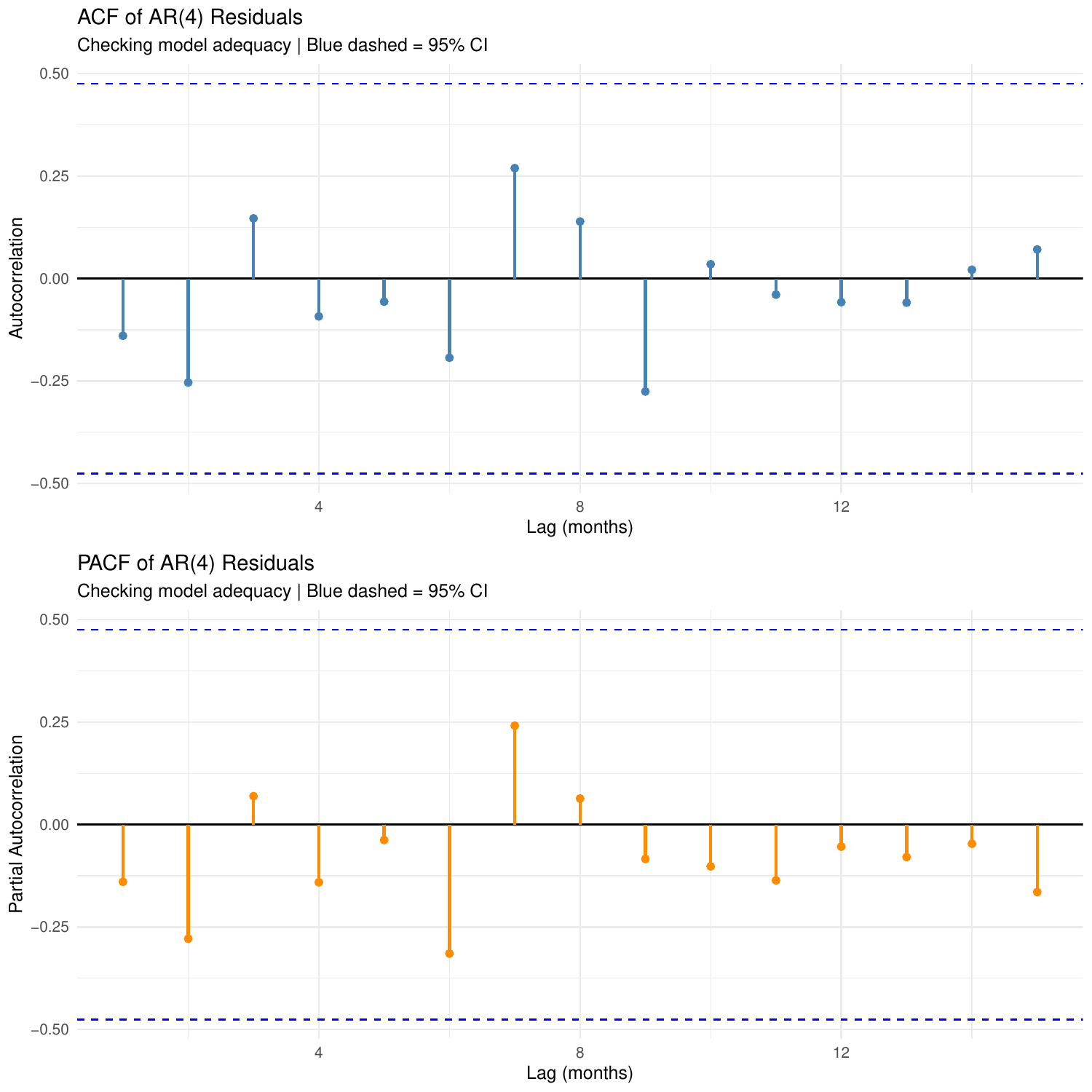}
    \caption{ACF and PACF of AR model residuals (post-fit). No significant autocorrelation remaining indicates model adequacy.}
    \label{fig:ar_resid_acf}
\end{figure}

To validate the AR model specification, we examine the autocorrelation function (ACF) and partial autocorrelation function (PACF) of the pre-merger detrended period FE residuals. Figure~\ref{fig:acf_pacf} displays the raw ACF/PACF, while Figure~\ref{fig:ar_resid_acf} shows the ACF/PACF of the fitted AR model residuals---the latter should show no significant autocorrelation if the model is adequate.

Table~\ref{tab:robust_acf} reports numerical ACF and PACF values along with the Ljung-Box test for residual autocorrelation.

\input{tex/APC_robust_table5_ACF_PACF.tex}

\section{Additional Heterogeneity Analyses}\label{app:het_additional}

\begin{figure}[h!]
    \centering
    \begin{subfigure}[b]{0.48\textwidth}
        \centering
        \includegraphics[width=\textwidth]{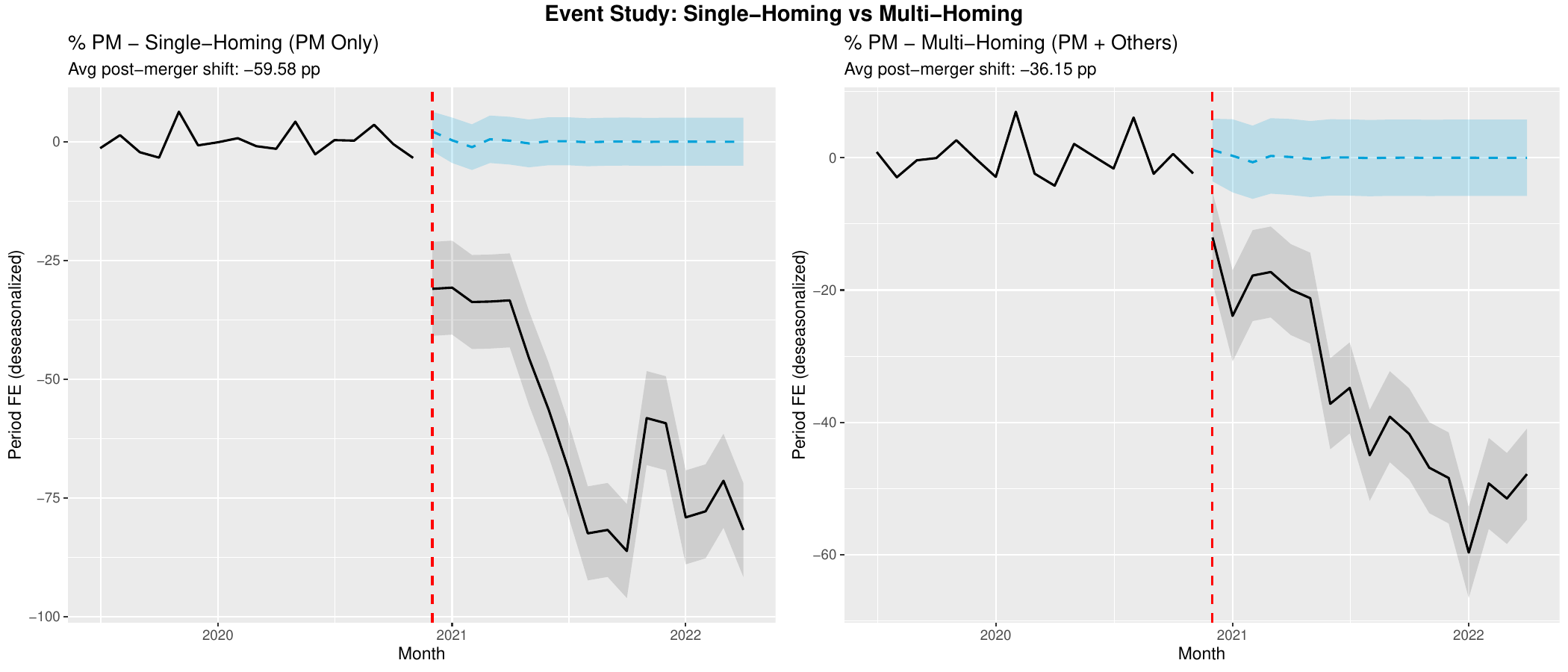}
        \caption{By multi-homing status}
    \end{subfigure}
    \hfill
    \begin{subfigure}[b]{0.48\textwidth}
        \centering
        \includegraphics[width=\textwidth]{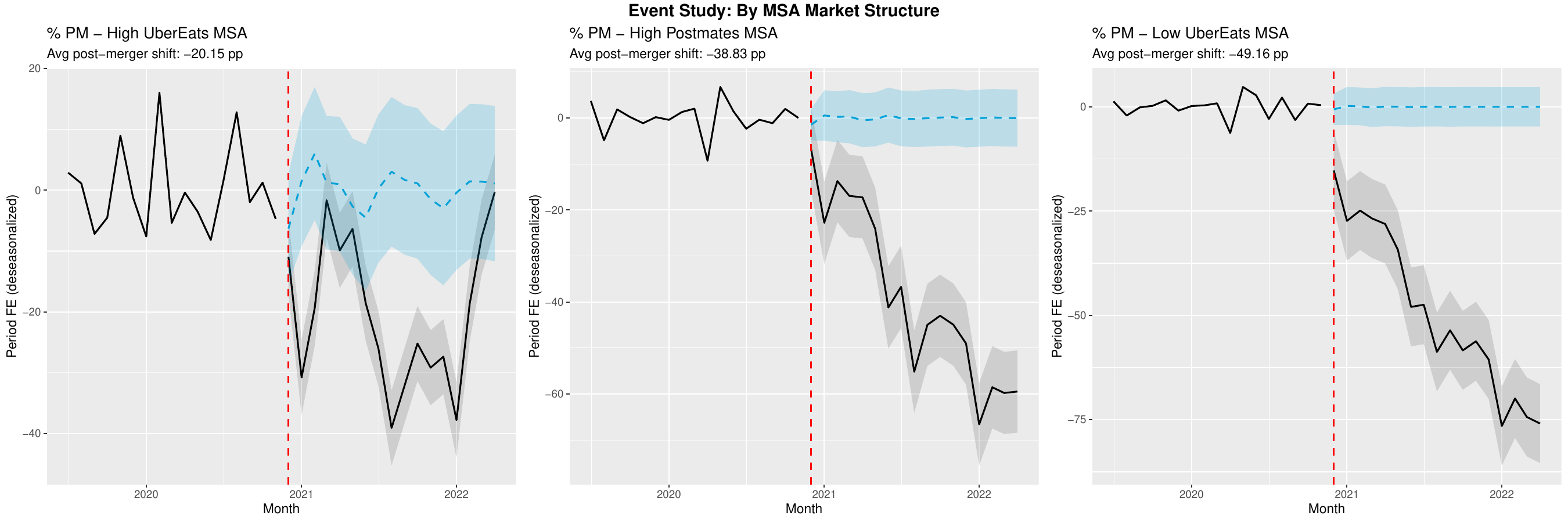}
        \caption{By MSA market structure}
    \end{subfigure}

    \vspace{0.5em}

    \begin{subfigure}[b]{0.48\textwidth}
        \centering
        \includegraphics[width=\textwidth]{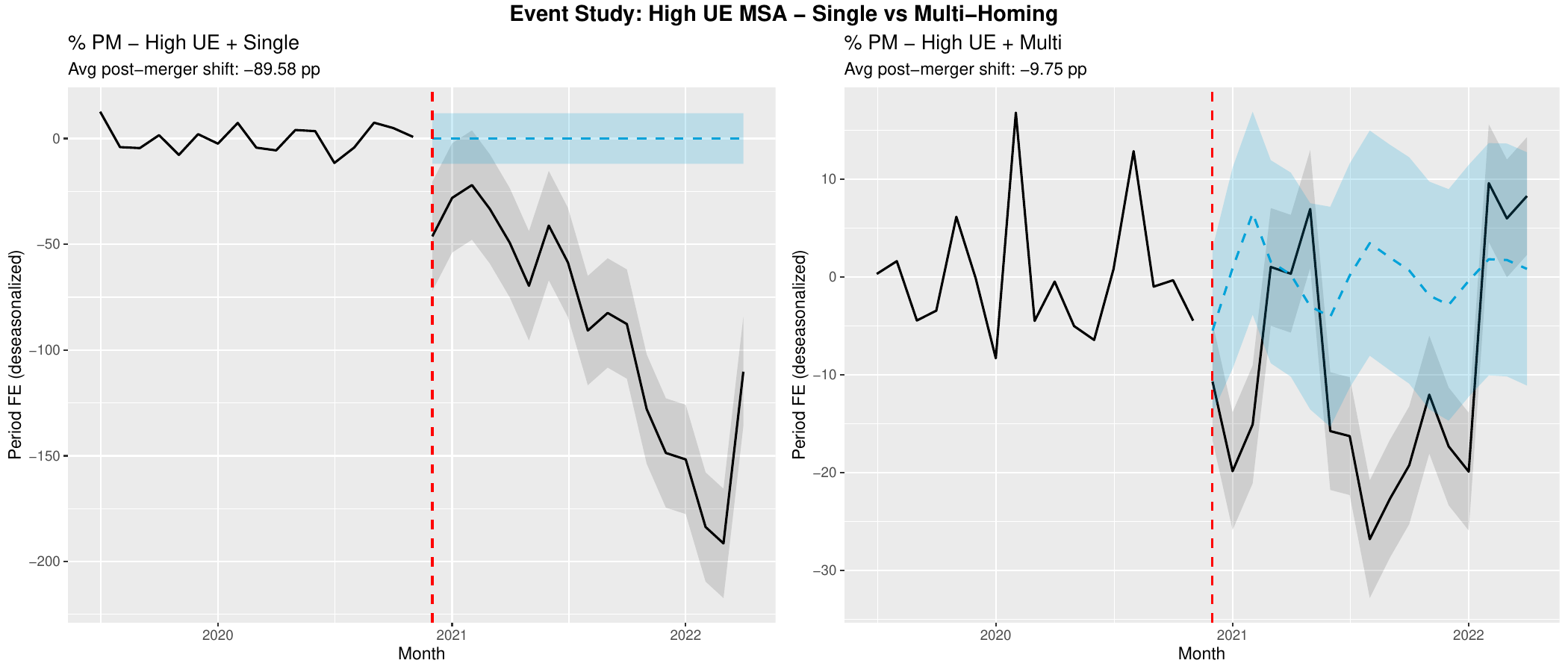}
        \caption{High UE MSA $\times$ multi-homing}
    \end{subfigure}
    \caption{Additional heterogeneity event studies. Each panel shows actual vs.\ counterfactual detrended period fixed effects with 95\% confidence intervals.}
    \label{fig:het_additional}
\end{figure}

Figure~\ref{fig:het_additional} presents event studies by multi-homing status, MSA market structure, and their interactions.

\section{Additional details for the DiD estimate}\label{app: did}

\subsection{Treatment Assignment}\label{subsec:treatment}

We exploit geographic variation in UberEats' pre-merger market presence to construct our treatment variable. For each MSA, we compute UberEats' pre-merger market share as the fraction of total food delivery spending (across UberEats, Postmates, DoorDash, and GrubHub) captured by UberEats:
\begin{equation}\label{eq:market_share}
    \text{UE\_Share}_{m} = \frac{\sum_{i \in m} \text{Spending}_{i,\text{UberEats},\text{pre}}}{\sum_{i \in m} \sum_{p \in \mathcal{P}} \text{Spending}_{i,p,\text{pre}}}
\end{equation}
where $m$ indexes MSAs, $i$ indexes consumers, $p$ indexes platforms, and $\mathcal{P} = \{\text{UberEats, Postmates, DoorDash, GrubHub}\}$ is the set of all platforms. The pre-merger period is defined as all months before December 2020.

MSAs are assigned to the treatment group if their UberEats market share exceeds the sample median:
\begin{equation}
    \text{Treated}_{m} = \mathbf{1}\left[\text{UE\_Share}_{m} > \text{Median}(\text{UE\_Share})\right]
\end{equation}

The intuition is that in MSAs where UberEats had a stronger pre-merger presence, the acquisition of Postmates would produce a larger competitive shock---effectively consolidating a greater share of the local market under a single firm. Consumers in high-UberEats-share MSAs are thus more ``treated'' by the merger.

\begin{figure}[H]
    \centering
    \includegraphics[width=0.95\textwidth]{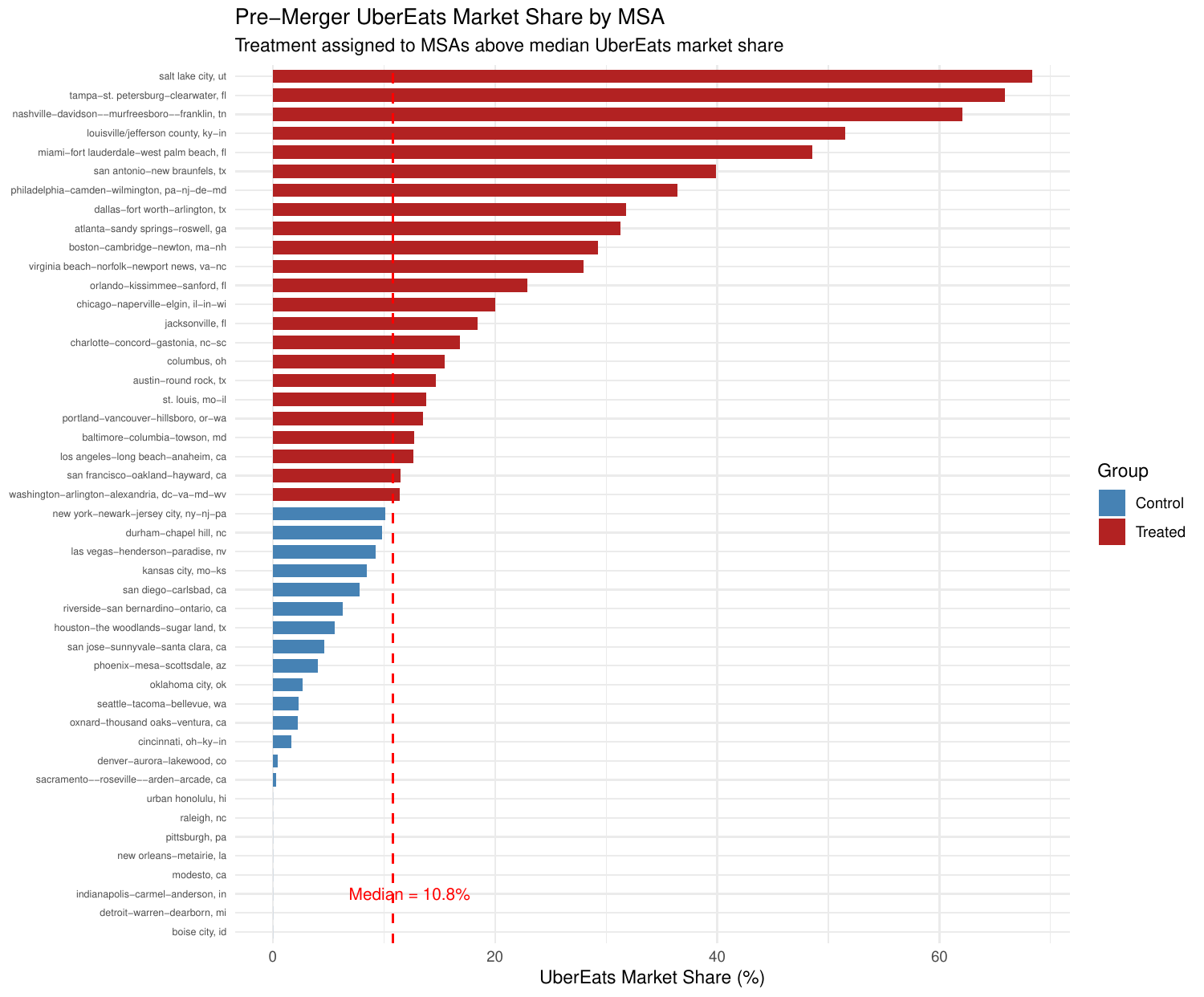}
    \caption{Pre-Merger UberEats Market Share by MSA. Treatment assigned to MSAs above the median UberEats market share (dashed red line).}
    \label{fig:ue_market_share}
\end{figure}
\subsection{Event Study Specification}\label{subsec:event_study}

To examine the dynamic effects of the merger and validate the parallel trends assumption, we estimate an event study specification for the DiD estimate :
\begin{equation}\label{eq:event_study}
    y_{imt} = \alpha + \sum_{\substack{\tau = \underline{\tau} \\ \tau \neq -1}}^{\bar{\tau}} \beta_{\tau} \cdot \mathbf{1}[t - t^{*} = \tau] \times \text{Treated}_{m} + \mathbf{X}'_{imt}\boldsymbol{\gamma} + \mu_{m} + \lambda_{\tau} + \varepsilon_{imt}
\end{equation}
where $\tau = t - t^{*}$ denotes months relative to the merger date ($t^{*} = \text{Dec 2020}$), and $\tau = -1$ (November 2020) serves as the reference period. Pre-merger coefficients $\{\beta_{\tau}\}_{\tau < 0}$ test the parallel trends assumption: if the identifying assumption holds, these coefficients should be statistically indistinguishable from zero.

We estimate this for both the full sample and the high-Postmates-share subgroup ($s_{i,\text{PM}} > 70\%$), with and without the full set of controls described above.

{Event Study Results}\label{subsec:event_results}

\begin{figure}[H]
    \centering
    \includegraphics[width=0.95\textwidth]{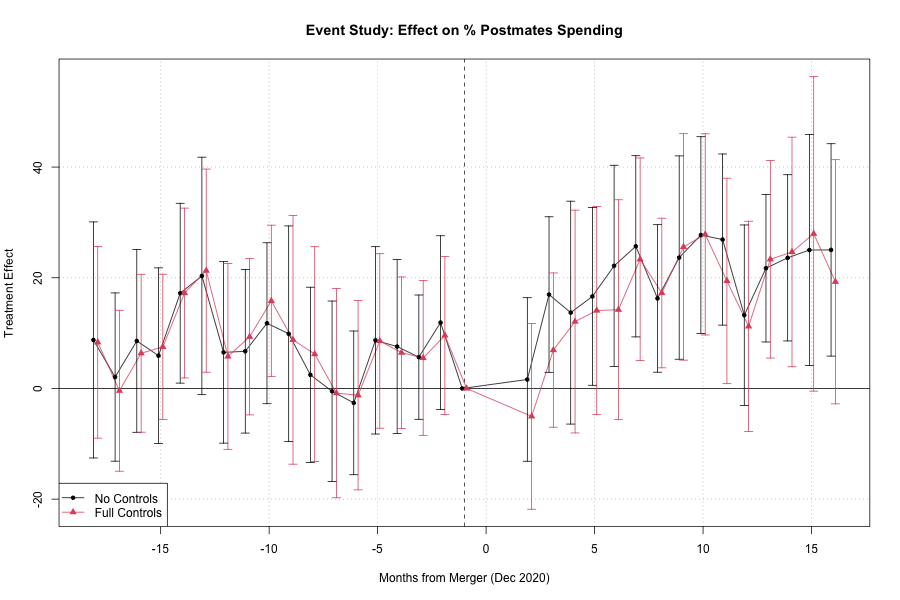}
    \caption{Event Study: Effect on Percentage Postmates Spending (All Users). Coefficients from Equation~\eqref{eq:event_study} with and without controls. Reference period is month $-1$ (November 2020).}
    \label{fig:event_study_all}
\end{figure}

\begin{figure}[H]
    \centering
    \includegraphics[width=0.95\textwidth]{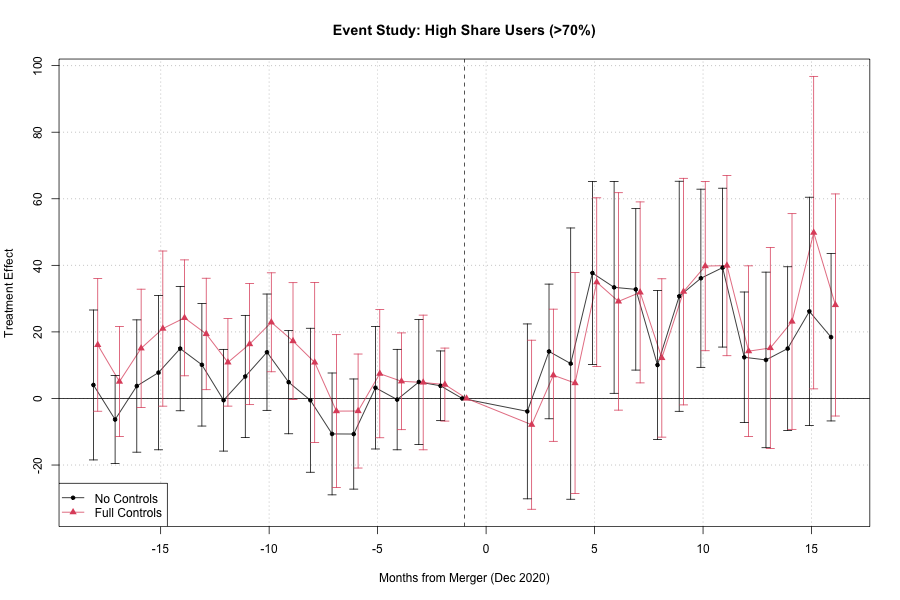}
    \caption{Event Study: High Postmates Share Subgroup ($>$70\% Pre-Merger Spending Share). Reference period is month $-1$ (November 2020).}
    \label{fig:event_study_high}
\end{figure}

\begin{figure}[H]
    \centering
    \includegraphics[width=0.95\textwidth]{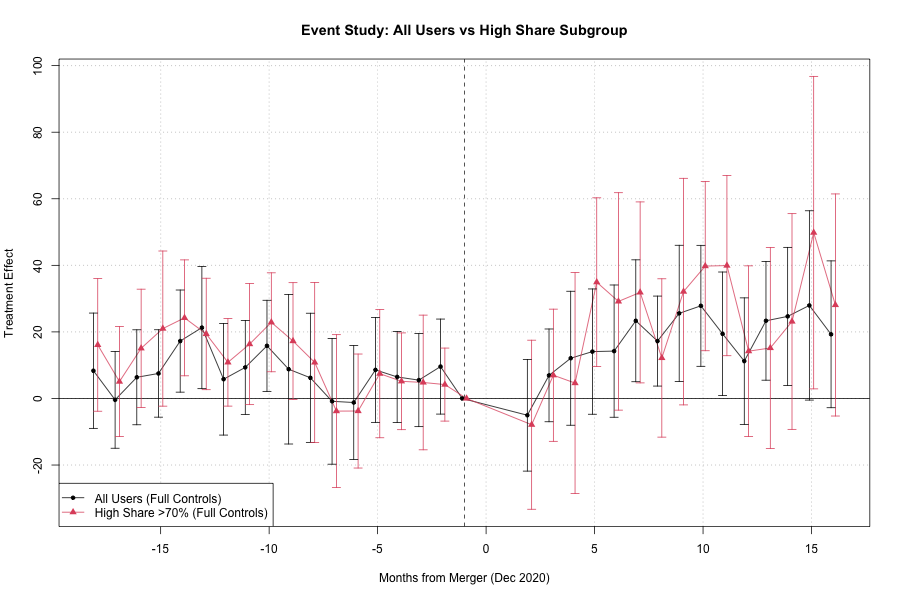}
    \caption{Event Study: All Users vs.\ High Share Subgroup (Full Controls). Comparison of dynamic treatment effects for the full sample and the high-dependence subgroup.}
    \label{fig:event_study_combined}
\end{figure}

Figures~\ref{fig:event_study_all}--\ref{fig:event_study_combined} present the event study estimates. Pre-merger coefficients that are close to zero and statistically insignificant would support the parallel trends assumption underlying our identification strategy.

\end{document}

%% file: tabs/receipts_summarystats.tex
\begin{tabular}{lcccc} \hline \hline
                    &    DoorDash&     Grubhub&   Postmates&    UberEats\\
\hline
Number of Active Users&       6,009&       3,423&       1,743&       4,303\\
Average Number of Orders Per User&          21&          12&           9&          18\\
Average Amount Paid per Order(\$)&       33.15&       35.06&       36.56&       31.15\\
Number of Orders    &     128,357&      42,600&      16,529&      79,098\\
\hline
\hline \end{tabular}

%% file: tex/APC_table2_OM_shifts.tex
\begin{table}[htbp]
\centering
\caption{Post-Merger Shift Estimates: All Users}
\label{tab:om_shifts}
\begin{tabular}{lcccc}
\toprule
 & \% Postmates & \% UberEats & \% DoorDash & \% GrubHub \\
\midrule
Post-merger shift & $-40.33^{***}$ & $+19.79^{***}$ & $+7.50^{***}$ & $+10.49^{***}$ \\
 & $(4.876)$ & $(3.299)$ & $(1.107)$ & $(1.270)$ \\
\midrule
COVID controls & \checkmark & \checkmark & \checkmark & \checkmark \\
User $\times$ time controls & \checkmark & \checkmark & \checkmark & \checkmark \\
Restaurant supply controls & \checkmark & \checkmark & \checkmark & \checkmark \\
Age FE & \checkmark & \checkmark & \checkmark & \checkmark \\
Cohort FE & \checkmark & \checkmark & \checkmark & \checkmark \\
Period FE & \checkmark & \checkmark & \checkmark & \checkmark \\
\bottomrule
\multicolumn{5}{l}{\footnotesize $^{***}p<0.01$, $^{**}p<0.05$, $^{*}p<0.10$. Standard errors of the mean post-merger lift in parentheses.} \\
\end{tabular}
\end{table}

%% file: tex/APC_table4_het_summary.tex
\begin{table}[htbp]
\centering
\caption{Heterogeneous Post-Merger Effects on \% Postmates Spending}
\label{tab:het_summary}
\begin{tabular}{lr}
\toprule
Subgroup &  Shift (pp) \\
\midrule
Low PM Budget Share ($ \leq 25\%$) & -14.82$^{***}$ \\
High PM Budget Share ($ \geq 70\%$) & -56.18$^{***}$ \\
Single-Homing & -59.58$^{***}$ \\
Multi-Homing & -36.15$^{***}$ \\
High PM Budget Share + Single & -59.58$^{***}$ \\
High PM Budget Share + Multi & -52.01$^{***}$ \\
High UE MSA & -20.15$^{***}$ \\
High PM MSA & -38.83$^{***}$ \\
Low UE MSA & -49.16$^{***}$ \\
High UE + Single & -89.58$^{***}$ \\
High UE + Multi & -9.75$^{**}$ \\
Low PM Budget Share+ High UE MSA & -4.04 \\
High PM Budget Share + High UE MSA & -43.07$^{***}$ \\
\bottomrule
\multicolumn{2}{l}{\scriptsize{$^{***}p<0.001$; $^{**}p<0.01$; $^{*}p<0.05$}} \\
\multicolumn{2}{l}{\scriptsize{\textit{Notes:} Avg gap between actual and ARIMA counterfactual period FEs;}} \\
\multicolumn{2}{l}{\scriptsize{ significance from one-sample $t$-test on post-merger lifts ($H_0$: lift $= 0$).}} \\
\end{tabular}
\end{table}

%% file: tex/APC_table6_OM_spec_curve_controls.tex
\begin{table}[htbp]
\centering
\caption{Post-Merger Shift in Period FEs Across Control Sets}
\label{tab:om_spec_curve}
\begin{tabular}{lc}
\hline\hline
Specification & Shift (pp) \\
\hline
1: No controls & -39.757 \\
2: + User (sync, multi) & -43.936 \\
3: + COVID & -59.564 \\
4: + Restaurants (full) & -40.139 \\
\hline\hline
\multicolumn{2}{l}{\footnotesize Shift = mean post-merger detrended FE $-$ mean pre-merger detrended FE.} \\
\multicolumn{2}{l}{\footnotesize Controls aggregated to cohort $\times$ month level (mean within group).} \\
\end{tabular}
\end{table}

%% file: tex/APC_robust_table1_AR_sensitivity.tex
\begin{table}[htbp]
\centering
\caption{Robustness: Sensitivity to AR Search Range}
\label{tab:robust_ar}
\begin{tabular}{lccccc}
\toprule
Search Range & Selected & AIC & Shift (pp) & 95\% CI Lo & 95\% CI Hi \\
\midrule
AR(0-3) & AR(0) & 119.3 & -40.139 & -55.081 & -25.197 \\
AR(0-6) & AR(4) & 115.9 & -40.333 & -53.072 & -27.594 \\
AR(0-9) & AR(4) & 115.9 & -40.333 & -53.072 & -27.594 \\
\bottomrule
\multicolumn{6}{l}{\footnotesize O\&M approach. Detrended with linear trend + bs(month, 3). Full controls.} \\
\end{tabular}
\end{table}

%% file: tex/APC_robust_table2_knot_sensitivity.tex
\begin{table}[htbp]
\centering
\caption{Robustness: Sensitivity to Seasonality Detrending (B-Spline Knots)}
\label{tab:robust_knots}
\begin{tabular}{lccccc}
\toprule
Knots & Selected AR & AIC & Shift (pp) & 95\% CI Lo & 95\% CI Hi \\
\midrule
3 & AR(4) & 115.9 & -40.333 & -53.072 & -27.594 \\
4 & AR(6) & 106 & -43.758 & -54.100 & -33.417 \\
5 & AR(6) & 104.9 & -42.366 & -52.282 & -32.451 \\
6 & AR(6) & 104.9 & -43.342 & -52.545 & -34.139 \\
\bottomrule
\multicolumn{6}{l}{\footnotesize O\&M approach. AR search 0--6. Full controls.} \\
\end{tabular}
\end{table}

%% file: tex/APC_robust_table3_controls_spec.tex
\begin{table}[htbp]
\centering
\caption{Robustness: O\&M Post-Merger Shift Across Control Sets}
\label{tab:robust_controls}
\begin{tabular}{lccccc}
\toprule
Specification & AR(p) & Shift (pp) & 95\% CI Lo & 95\% CI Hi \\
\midrule
1: No controls & AR(6) & -39.681 & -44.603 & -34.758 \\
2: + User (sync/multi) & AR(6) & -43.876 & -48.944 & -38.809 \\
3: + COVID & AR(6) & -59.736 & -69.959 & -49.514 \\
4: + Restaurants (full) & AR(4) & -40.333 & -53.072 & -27.594 \\
\bottomrule
\multicolumn{5}{l}{\footnotesize Controls added cumulatively. bs(month, 3). AR search 0--6.} \\
\end{tabular}
\end{table}

%% file: tex/APC_robust_table5_ACF_PACF.tex
\begin{table}[H]
\centering
\caption{ACF and PACF of Pre-Merger Detrended Period FE Residuals}
\label{tab:robust_acf}
\begin{tabular}{ccccc}
\toprule
Lag & ACF & Significant? & PACF & Significant? \\
\midrule
1 & 0.089 &  & 0.089 &  \\
2 & -0.407 &  & -0.418 &  \\
3 & -0.193 &  & -0.127 &  \\
4 & -0.342 &  & -0.597 & Yes \\
5 & 0.012 &  & -0.181 &  \\
6 & 0.343 &  & -0.291 &  \\
7 & 0.291 &  & 0.121 &  \\
8 & -0.052 &  & -0.305 &  \\
9 & -0.275 &  & -0.089 &  \\
10 & -0.077 &  & -0.096 &  \\
11 & -0.007 &  & 0.029 &  \\
12 & 0.057 &  & -0.080 &  \\
\midrule
\multicolumn{5}{l}{Ljung-Box Q(12) = 10.12, $p$ = 0.6052} \\
\bottomrule
\multicolumn{5}{l}{\footnotesize 95\% CI = $\pm$ 0.475. n = 17 pre-merger months.} \\
\end{tabular}
\end{table}